\documentclass[11pt,a4paper]{article}
\usepackage{jheppub}  
\usepackage{slashed,xcolor}
\usepackage[normalem]{ulem}
\boldmath
\title{\boldmath
Structure functions for the inclusive semileptonic $b$-quark decay at NNLO: a semi-analytic calculation \unboldmath}

\author[a]{A. Broggio,}
\author[b]{B. Capdevila,}
\author[c]{A. Ferroglia,}
\author[d]{P. Gambino}
\affiliation[a]{Faculty of Physics, University of Vienna, Boltzmanngasse 5, A-1090 Vienna, Austria}
\affiliation[b]{Departament de Física Quàntica i Astrofísica, Institut de Ciències del Cosmos (ICCUB), Universitat de Barcelona, Martí Franquès 1, E08028 Barcelona}
\affiliation[c]{Physics Department, New York City College of Technology, The City University of New York, 300 Jay Street, Brooklyn, NY 11201, USA}
\affiliation[d]{Dipartimento di Fisica, Universit\`a di Torino \& INFN, Sezione di Torino\\
via Giuria 1, 10124 Torino,
Italy}

\emailAdd{alessandro.broggio@univie.ac.at}
\emailAdd{bcapdevila@icc.ub.edu}
\emailAdd{andrea.ferroglia@gmail.com}
\emailAdd{paolo.gambino@unito.it}

\abstract{ \unboldmath
We present a study of the inclusive charmless semileptonic $b$ decay, $b\to X_u\ell\bar{\nu}$ at next-to-next-to-leading order (NNLO) in perturbative QCD, with the primary aim of extracting the hadronic structure functions $W_i$ at NNLO. The analysis is based on a numerical calculation of the relevant kinematic distributions using a phase-space slicing method to handle infrared-sensitive contributions from real gluon emissions. We use known results from Heavy Quark Effective Theory and Soft-Collinear Effective Theory to extract the singular terms and construct a model for the regular contributions to the structure functions at NNLO, then perform a fit to the numerical results. We use our approximate  structure functions to compute various kinematic distributions and moments: the comparison with existing analytic and numerical results shows very good agreement, which is further improved including the available analytic results in the fit.
}

\begin{document}
\begin{flushright}
UWThPh 2026-1
\end{flushright}
\maketitle
\flushbottom

\newcommand{\mxs}{\ensuremath{\hat{m}_X^2}}
\def \be{\begin{equation}}
\def \ee{\end{equation}}

\section{Introduction}
\label{sec:intro}
Precise theoretical predictions for the inclusive charmless semileptonic decay $B \to X_u\ell\nu$ are essential for determining the magnitude of the Cabibbo–Kobayashi–Maskawa (CKM) matrix element $V_{ub}$. As stringent phase-space cuts must be generally employed to suppress the dominant $B\to X_c \ell \nu$ background, the theoretical description  of this inclusive  decay is based on a non-local Operator Product Expansion (OPE) \cite{Neubert:1993ch,Bigi:1993ex}, where nonperturbative shape functions (SFs) play a role akin to that of parton distribution functions of the $b$ quark inside the $B$ meson.

Among the theoretical frameworks that incorporate this formalism, Bosch-Lange -Neubert-Paz (BLNP) \cite{Lange:2005yw}, Gambino-Giordano-Ossola-Uraltsev (GGOU) \cite{Gambino:2007rp}, and  Dressed Gluon Exponentiation (DGE) \cite{Andersen:2005mj} are currently employed by the Heavy Flavour Averaging Group (HFLAV)  \cite{hflav2024}. The latest average values of $|V_{ub}|$ in these three
frameworks are
\begin{equation}
\label{averages}
|V_{ub}|^{\rm BLNP}\!\!=4.25(23) \times 10^{-3}\,,\quad
|V_{ub}|^{\rm GGOU}\!\!=4.06(16) \times 10^{-3}\,,\quad
|V_{ub}|^{\rm DGE}\!\!=3.87(14) \times 10^{-3}\, ,
\end{equation}
in reasonable agreement with each other\footnote{The averages in (\ref{averages}) do not include the new results by Belle II \cite{Belle-II:2025pye}. They are compatible with (\ref{averages}) but generally prefer a lower value of $|V_{ub}|$.}. However, the values obtained from different experimental analyses are not always compatible within their stated theoretical and experimental uncertainties. In particular,  the single most precise value in the above averages corresponds to the 2016 endpoint analysis by BaBar \cite{BaBar:2016rxh}, which shows a strong dependence on the model used to simulate the signal and leads to sharply different results in BLNP and GGOU. Lower values of  $|V_{ub}|$ are in better agreement with
\be
|V_{ub}|^{B\to \pi\ell\nu}= 3.75(20) \times 10^{-3}\, , \label{eq:1}
\ee
the value extracted from $B\to \pi \ell\nu$ data together with lattice QCD determinations of the relevant form factor \cite{hflav2024}.

The Belle II experiment at KEK should help clarify  the matter in various ways,  see \cite{Gambino:2020jvv}. In particular, it should be possible to calibrate and validate the different frameworks directly on data, especially on differential distributions which are sensitive to the SFs, improving on the first measurement of such differential distributions by Belle  \cite{Belle:2021ymg}. The SIMBA \cite{Ligeti:2008ac} and  NNVub \cite{Gambino:2016fdy} methods both aim at a model-independent parametrization of the relevant SFs and are well posed to analyze the future Belle II data in an efficient way.

In view of these interesting prospects, various improvements are necessary on the theoretical side, among which the inclusion of $O(\alpha_s^2)$ corrections to the underlying partonic decay not enhanced by $\beta_0$ and of $O(\alpha_s/m_b^2)$ effects that modify the OPE constraints on the SFs, see \cite{Capdevila:2021vkf}.

At next-to-next-to-leading order (NNLO) in QCD ($\mathcal{O}(\alpha_s^2)$), we have analytic results for the total width of the partonic process $b\to X_u\ell\bar{\nu}$ \cite{vanRitbergen:1999gs}, for a few moments \cite{Pak:2008qt}, and for the $q^2$ differential distribution \cite{Czarnecki:2001cz,Chen:2022wit}, as well as a numerical calculation of various moments \cite{Brucherseifer:2013cu}. Partial results are known even for the total width at NNNLO \cite{Fael:2023tcv}. However, the NNLO implementation of the theoretical frameworks requires the structure functions  $W_i$ that decompose the hadronic tensor as a starting point, and the analytic calculation of the  partonic $b\to X_u\ell\bar{\nu}$ triple differential  decay distribution (equivalent to that of the structure functions) is still incomplete in the full phase space. While all the $O(\alpha_s^2)$ corrections  enhanced by $\beta_0$ (or BLM \cite{Brodsky:1982gc}) have been computed in \cite{Gambino:2006wk}, the situation is different for those not enhanced by $\beta_0$: the virtual corrections were computed  in 2008 by several groups \cite{Bonciani:2008wf,Asatrian:2008uk,Beneke:2008ei,Bell:2008ws}, but the real contributions are not fully known. A few years ago, a partial result for the double real-radiation corrections was obtained~\cite{Bonciani:2018fvk}. However, even for the double emission contribution a single non-planar topology remains uncomputed and,  in addition, the mixed real–virtual (one-real–one-virtual) contributions are not yet evaluated. As a result, the analytic form of the NNLO real-emission correction to $b\to X_u\ell\bar{\nu}$ is not yet known.  

In this paper, we discuss the semi-analytic extraction of the hadronic form factors $W_i$ up to NNLO. The method which we employ to obtain this semi-analytic result is based on three elements:
\begin{itemize}
\item[i)] A parton-level Monte Carlo code for the evaluation of differential distributions of the partonic decay $b\to X_u\ell\bar{\nu}$,
\item[ii)] the analytic expression of the singular terms (i.e.\ terms proportional to plus distributions and to the Dirac delta function) in the NNLO form factors~\cite{Bosch:2004th,Becher:2006qw,Asatrian:2008uk}, and 
\item[iii)] the analytic expression of the BLM $\mathcal{O}(\alpha_s^2)$ corrections to the form factors \cite{Gambino:2006wk}.
\end{itemize}
By combining these three elements, we obtain expressions for the hadronic structure functions $W_i$ at $\mathcal{O}(\alpha_s^2)$ where the terms which are not singular or part of the BLM corrections are described by a precise numerical fit, which can be further improved by including the available analytic $\mathcal{O}(\alpha_s^2)$  results for the total width, the first two total lepton energy moments and the leptonic invariant mass ($q^2$) distribution \cite{vanRitbergen:1999gs,Chen:2022wit,Pak:2008qt}. We refer to these results as the semi-analytic approximation of the form factors $W_i$ at NNLO. 

We begin in Section~\ref{sec:InclusiveNLO} by reviewing the kinematics and the NLO results for inclusive $b \to u$ semileptonic decays. In Section~\ref{sec:SCET}, we discuss the structure of the singular terms contributing to the NNLO form factors. These terms can be obtained by means of Effective Field Theory (EFT) methods based on the Heavy Quark Effective Theory (HQET) and Soft-Collinear Effective Theory (SCET). 
Section~\ref{sec:slicing} describes the partonic Monte Carlo code employed to evaluate numerically  differential distributions and observables for $b\to X_u\ell\bar{\nu}$ decays up to NNLO. In particular, we discuss the implementation of the slicing method used in the code, including the treatment of infrared divergences in the real-emission contribution. This method partitions the phase space using an unphysical cut parameter to separate unresolved infrared regions from hard-emission regions; we explain how this is applied to the decay $b\to X_u\ell\bar{\nu}$ and assess the associated systematic uncertainties.
In Section~\ref{sec:form-factors}, we present our methodology for extracting the $W_i$ structure functions; we describe how the two-loop singular and regular contributions to the form factors are combined and fitted to the numerical results obtained by the slicing method.  We detail the evaluation of the phase-space integrals, including key ``preprocessing'' integrals used to match to the binned numerical results from the slicing method. 

In Section~\ref{sec:NLO} we validate our approach by comparing results obtained with our method, applied to NLO, to the known analytic results at NLO. We illustrate the outcome of this comparison by considering several distributions depending on a single kinematic variable. In particular, we present the leptonic invariant mass $q^2$, the hadronic invariant mass $m_X^2$, and the charged-lepton energy $E_\ell$ distributions at NLO. These serve as a baseline for the NNLO analysis. In Section~\ref{sec:NNLO} we discuss the functional basis employed to fit the NNLO form factors $W_i$.  

Section~\ref{sec:results} provides results for the $q^2$, $m_X^2$, and $E_\ell$ spectra and total moments at NNLO. In each case, we compare the results obtained starting from our  approximate expressions for the form factors to existing analytic results in the literature and to the analogous results that can be computed from the parton-level Monte Carlo. The agreement of the distributions based on the semi-analytic $W_i$ with the corresponding numerical and analytic results demonstrates the internal consistency and robustness of our NNLO extraction. Finally, in Section~\ref{sec:conclusions}, we summarize our results.

\section{Inclusive decays and NLO contributions}
\label{sec:InclusiveNLO}

We consider the decay of  an on-shell $b$ quark with four-momentum $p = m_b v$ into a lepton pair with momentum $q = p_\ell + p_{\nu_\ell}$ and a hadronic final state with momentum $p^\prime = p - q$ in perturbative QCD. Assuming the hadronic final state contains a massless up quark, we express the $b$-quark decay kinematics in terms of the dimensionless variables\footnote{It is worth noting that in other works, such as Refs.~\cite{Alberti:2012dn,Capdevila:2021vkf}, the symbol $\hat{u}$ is employed instead of $\mxs$.}
\begin{equation} \label{eq:variables}
\mxs = \frac{(p - q)^2}{m_b^2}, \qquad \hat{q}_0 = \dfrac{q_0}{m_b} = \dfrac{v\cdot q}{m_b}, \qquad \hat{q}^2 = \frac{q^2}{m_b^2}\, .
\end{equation}
The physical range of these variables is
\begin{equation}
0 \le \mxs \le (\mxs)_+ = (1 - \sqrt{\hat{q}^2})^2\,, \qquad \sqrt{\hat{q}^2} \le \hat{q}_0 \le \dfrac{1 + \hat{q}^2}{2}\,, \qquad \text{and} \qquad 0 \le \hat{q}^2 \le 1\, .
\label{eq:physrange}
\end{equation}
The equations above, together with eq.~\eqref{eq:elhatregion} for the charged lepton energy $E_\ell = m_b \hat{E}_\ell$ describe the complete real emission phase space.
The variables $\hat{q}_0$, $\hat{q}^2$ and $\mxs$ are not independent but satisfy the relation
\begin{equation}
\mxs =  1 +\hat{q}^2 - 2 \hat{q}_0 \, .
\end{equation}
We also introduce the variable
\begin{equation}
w = 1 - \hat{q}^2\, ,
\end{equation}
and the hadronic energy normalised to the $b$-quark mass:
\begin{equation}
E = \frac{1}{2} (w + \mxs)\, .
\end{equation}
The case of tree-level kinematics corresponds to $\mxs = 0$, with the corresponding hadronic final-state energy given by
\begin{equation}
E_0 = \frac{1}{2} w\, .
\end{equation}
The normalised total leptonic energy is
\begin{equation}
\hat{q}_0 = 1 - E\, , \qquad \text{from which it follows} \qquad \mxs = 2(1 - E_0 - \hat{q}_0)\, .
\end{equation}
We further define a threshold factor
\begin{equation}
\lambda = 4(\hat{q}_0^2 - \hat{q}^2) = 4(E^2 - \mxs).\label{eq:lambda}
\end{equation}
In the case of tree-level kinematics, the threshold factor becomes $\lambda_0 = 4 E_0^2$.

The fully differential decay rate can be expressed as 
\begin{equation}
\frac{d\Gamma}{dq^2\, dq_0\, dE_\ell} = \frac{G_F^2 |V_{ub}|^2}{8\pi^3}\,
L_{\mu\nu}(p_\ell, q)\, W^{\mu\nu}(p, q)\, ,
\end{equation}
where $L_{\mu\nu}$ is the leptonic tensor
\begin{equation}
L^{\mu\nu} = 2\, (p_\ell^\mu p_{\nu_\ell}^\nu + p_\ell^\nu p_{\nu_\ell}^\mu - g^{\mu\nu}p_\ell p_{\nu_\ell} + i \epsilon^{\mu\nu\rho\sigma} p_{\ell\rho}p_{{\nu_\ell}\sigma})\, .
\end{equation}
Parity conservation and Lorentz symmetry allow for a decomposition of $W^{\mu\nu}$ in terms of scalar structure functions $W_i$:
\begin{equation}
m_b\, W^{\mu\nu}(p, q) = -W_1\, g^{\mu\nu} + W_2\, v^\mu v^\nu + i W_3\, \epsilon^{\mu\nu\rho\sigma} v_\rho \hat{q}_\sigma + W_4\, \hat{q}^\mu \hat{q}^\nu + W_5 \left( v^\mu \hat{q}^\nu + v^\nu \hat{q}^\mu \right)\, ,
\label{eq:Wmunu-decomp}
\end{equation}
where $\hat{q}^\mu = q^\mu /m_b$ .
The arguments of the scalar structure functions $W_i$ can be taken as $q_0$ and $q^2$. 

Upon contracting $L_{\mu\nu}$ with the decomposition in eq.~\eqref{eq:Wmunu-decomp}, one finds that in the massless leptons case only $W_1$, $W_2$, and $W_3$ contribute:
\begin{align}
\frac{d \Gamma}{\, d\hat{q}^2 \, d\hat{q}_0\, d\hat{E}_\ell} & = 
\frac{G_F^2 m_b^5 |V_{ub}|^2}{8\pi^3} \, 
\left\{
\hat{q}^2 W_1 - \left[ 2 \hat{E}_\ell^2 - 2 \hat{E}_\ell \hat{q}_0 + \frac{\hat{q}^2}{2} \right] W_2 + \hat{q}^2 (2 \hat{E}_\ell - \hat{q}_0) W_3 \right\},
\label{eq:triple}
\end{align}
where we neglected the theta functions related to the phase-space boundaries. 

If one integrates over the lepton energy $\hat{E}_l$ in the allowed range
\begin{equation}\label{eq:elhatregion}
    \frac{1}{2} \left( \hat{q}_0 - \sqrt{\hat{q}_0^2 - \hat{q}^2} \right) \le \hat{E}_l \le \frac{1}{2} \left( \hat{q}_0 + \sqrt{\hat{q}_0^2 - \hat{q}^2} \right) \, ,
\end{equation}
and trades $\hat{q}_0$ for $\mxs$ through the relation $\hat{q}_0  = (1+\hat{q}^2 -\mxs)/2$,
it is possible to obtain the double differential decay rate
\begin{eqnarray}
\frac{d \Gamma}{d\hat q^2 \, d \mxs }&=&
\frac{G_F^2 m_b^5 |V_{ub}|^2}{16\pi^3}
\sqrt{\hat q_0^2-\hat q^2}\left\{
\hat q^2\, W_1  +\frac13 (\hat q_0^2-\hat q^2) W_2 \right\}\, .
\label{eq:double}
\end{eqnarray}
By repeating the same integration with a weight function $E_\ell$ and $E_\ell^2$ one obtains the first and second (double-differential) lepton energy moments as a function of the form factors $W_i$:
\begin{eqnarray}
\frac{d \langle E_\ell \rangle}{d\hat q^2 \, d \mxs }&=&
\frac{G_F^2 m_b^5 |V_{ub}|^2}{32\pi^3}
\sqrt{\hat q_0^2-\hat q^2}\Biggl\{
\hat q^2 \hat q_0\, W_1  +\frac13 \hat q_0(\hat q_0^2-\hat q^2) W_2 
\nonumber \\
& & \hspace{3.9cm} 
+\frac13 \hat q^2(\hat q_0^2-\hat q^2) W_3\Biggr\} \, ,
\label{eq:double1}\\
\frac{d \langle E^2_\ell \rangle}{d\hat q^2 \, d \mxs }&=&
\frac{G_F^2 m_b^5 |V_{ub}|^2}{96\pi^3}
\sqrt{\hat q_0^2-\hat q^2}\Biggl\{
\frac{\hat q^2}2 (4\hat q_0^2-\hat q^2)\, W_1  +\frac{6\hat q_0^2-\hat q^2}{10}(\hat q_0^2-\hat q^2) W_2 \nonumber \\ 
& & \hspace{3.9cm} +\hat q_0 \hat q^2(\hat q_0^2-\hat q^2) W_3\Biggr\} \, .
\label{eq:double2}
\end{eqnarray}
We now expand the form factors in powers of $\alpha_s$
\begin{equation} \label{eq:Wexp}
W_i (\hat{q}_0,\hat{q}^2) = W_i^{(0)} (\hat{q}_0,\hat{q}^2) +
C_F \frac{\alpha_s(\mu)}{\pi} W_i^{(1)}(\hat{q}_0,\hat{q}^2) + 
C_F \left( \frac{\alpha_s(\mu)}{\pi} \right)^2 W_i^{(2)}(\hat{q}_0,\hat{q}^2,\mu) + O(\alpha_s^3)\, ,
\end{equation}
In the Standard Model, the leading-order coefficients are
\begin{equation}
W_i^{(0)} = w_i^{(0)}\, \delta(\mxs)\, , \qquad
w_1^{(0)} = 1 - \hat{q}^2 \, , 
\qquad w_2^{(0)} = 4, \qquad w_3^{(0)} = 2\, .
\end{equation}
At $\mathcal{O}(\alpha_s)$, one finds~\cite{DeFazio:1999ptt,Capdevila:2021vkf}
\begin{align}
W_i^{(1)} =& w_i^{(0)} \left\{ \mathcal{S}_i\, \delta(\mxs) 
- \left[ \frac{\ln \mxs}{\mxs} \right]_+ 
- \left( \frac{7}{4} - 2 \ln w \right) \left[ \frac{1}{\mxs} \right]_+
+ w\, B(\hat{q}^2, \mxs)\, \theta(\mxs) \right\} \nonumber \\
&+ \mathcal{R}_i^{(1)}\, \theta(\mxs)\, ,
\label{NLO}
\end{align}
where
\begin{equation}
\mathcal{S}_i = -\frac{5}{4} - \frac{\pi^2}{3} - \text{Li}_2(1 - w)
- 2 \ln^2 w - \frac{5w - 4}{2(1 - w)} \ln w
+ \frac{\ln w}{2(1 - w)}\, \delta_{i2}\, , 
\label{eq:Si}
\end{equation}
and
\begin{equation}
B(\hat{q}^2, \mxs) = \frac{\ln (\mxs/w^2) + w \mathcal{I}_1}{w \mxs} 
\simeq \frac{w - 2}{w^3} \ln \frac{\mxs}{w^2} 
+ \frac{2(w - 1)}{w^3} + \mathcal{O}(\mxs)\, ,
\label{eq:Bexp}
\end{equation}
with
\begin{equation} \label{eq:I_1}
\mathcal{I}_1 = \frac{1}{\sqrt{\lambda}} \ln \frac{\mxs + w + \sqrt{\lambda}}{\mxs + w - \sqrt{\lambda}}\,.
\end{equation}
The functions $\mathcal{R}_i^{(1)}$ are given by:
\begin{align}
\mathcal{R}_1^{(1)} &= \frac{3}{4} + \frac{\mxs (12 - w - \mxs)}{2 \lambda}
+ \left( w + \frac{\mxs}{2} - \frac{\mxs(2 \mxs + 3 w)}{\lambda} \right) \mathcal{I}_1\,,\nonumber\\
\mathcal{R}_2^{(1)} &= \frac{6 \mxs (\hat{m}_X^4 - (3 - w) \mxs - 12 + 13 w)}{\lambda^2}
+ \frac{\mxs - 38 + 21 w}{\lambda} \nonumber \\
& \quad - 4 \frac{ \frac{w}{2} \hat{m}_X^6 + (2 w^2 - 6) \hat{m}_X^4 + (7 - 3w + \frac{5}{2} w^2) w \mxs + w^3 (w - 4)}{\lambda^2} \mathcal{I}_1\, ,\nonumber\\
\mathcal{R}_3^{(1)} &= \frac{3 \mxs - 8 + 5 w}{\lambda} 
+ \frac{ \hat{m}_X^4 - (6 - w) \mxs + 4 w }{ \lambda } \mathcal{I}_1 \, .
\end{align}
The goal of this work is to obtain semi-analytic expressions for the NNLO form factors $W^{(2)}_i$.

\section{Effective Field Theory Structures at NNLO}
\label{sec:SCET}

To introduce the theoretical framework required for our analysis, we begin by considering the decay of a $B$-meson into a $u$-flavored inclusive hadronic state $X_u$ and a dilepton: $B\to X_u\ell\nu$. In the kinematic region where the hadronic final states are constrained to have large energy $E_X \sim m_B$ but only moderate invariant mass $m_X \sim \sqrt{m_B \Lambda_\mathrm{QCD}} \ll m_B$, an  EFT  approach to the decay process is particularly suitable. In this kinematic limit, called the ``shape-function region'', the decay rates and spectra are obtained using a twist expansion involving non-perturbative functions named \textit{shape functions}~\cite{Neubert:1993um, Bigi:1993ex, Mannel:1994pm}, which describes the non-perturbative internal structure of the $B$-meson. At the same time, the separation of the scales characterizing the physical process,
\begin{displaymath}
m_B\;\mathrm{(hard)}, \qquad \sqrt{m_B\Lambda_\mathrm{QCD}}\;\mathrm{(hard-collinear)}, \qquad \Lambda_\mathrm{QCD}\;\mathrm{(soft)},
\end{displaymath}
makes the calculation of (short-distance) perturbative corrections more complicated than in most applications of HQET \cite{Neubert:1993mb,Manohar:2000dt}. Logarithms involving the different scales appear at each order of the perturbative expansion, and they can be appropriately resummed~\cite{Falk:1993vb, Bosch:2004th}.

In order to resum these large logarithms, one can employ an  EFT framework which combines HQET and  SCET. A systematic treatment of this problem within this framework was developed in \cite{Bosch:2004th}. The main result of the matching of the EFT to QCD is a factorization formula for the differential distribution
\begin{equation}\label{eq:erSCETFactorisation}
d\Gamma \sim H\cdot J \otimes S + {\mathcal O}\left(\dfrac{\Lambda_{\mathrm{QCD}}}{m_b}\right)+\ldots \, ,
\end{equation}
where power corrections suppressed by the $b$-quark mass or $\mxs$ \footnote{The power suppression is relative to the leading power contributions, therefore singular but integrable terms proportional to $\ln^n(\mxs)$ are neglected.} are neglected and $\otimes$ indicates a convolution. $H$, $J$ and $S$ are single-scale functions that depend only on the hard, collinear and soft degrees of freedom, respectively.

Naturally, the factorization in hard, soft, and collinear factors can be applied directly to the hadronic tensor 
\begin{equation}\label{eq:erSCETFactorisationWi}
W^{\mu\nu} = \sum_{i,j=1}^3 H_{ij}(\bar{n}\cdot p)\,\frac{(\bar{n}\cdot p)}{2}\,\mathrm{tr}\left(\bar{\Gamma}_j^\mu\,\dfrac{\slashed{n}}{2}\,\Gamma_i^\nu\,\dfrac{1 + \slashed{v}}{2}\right)\int d\omega\, J(p_\omega^2)S(\omega) + \ldots\,  ,
\end{equation}
where $\Gamma_i$ are the following Dirac structures \cite{Asatrian:2008uk}
\begin{align}
\Gamma_1^\mu=\gamma^\mu (1-\gamma_5)\, ,\qquad \Gamma_2^\mu=v^\mu (1+\gamma_5)\, , \qquad \Gamma_3^\mu=\frac{n^\mu}{n\cdot v}(1+\gamma_5)\,\, .
\end{align}
The two light-like vectors  $n^\mu$ and $\bar{n}^\mu$ introduced above satisfy the relations $n^2=0$, $\bar{n}^2=0$, and $\bar{n}\cdot n = 2$. In addition,  we defined $p_\omega^\mu = p^\mu + \frac{1}{2}\omega\bar{n}^\mu$. The structures involved in the convolution in eq.~\eqref{eq:erSCETFactorisation} encapsulate the separation of scales achieved through the EFT description of the decay:
\begin{itemize}
\item[\textit{i})] The \textit{hard} functions $H_{ij}$ are perturbatively computable structures that contain the information on the physics of the decay process at the hard scale $m_b$. The $H_{ij}$ functions are obtained from integrating out the hard scale $m_b$ by matching the $b \to u$ transition current from full QCD onto SCET. More precisely,
\begin{equation} \label{eq:hardfunc}
H_{ij}(\bar{n}\cdot p) = C_j^*(\bar{n}\cdot p)\,C_i(\bar{n}\cdot p)\, ,
\end{equation}
where the functions $C_i$ are the Fourier-transformed Wilson coefficients of the effective theory. The $C_i$ matching coefficients are currently known at NLO~\cite{Bosch:2004th} and NNLO in QCD~\cite{Asatrian:2008uk}.
\item[\textit{ii})] The \textit{jet} function $J$ is also a short-distance object that can be calculated in perturbation theory. This function and its QCD corrections can be computed by evaluating the discontinuity of the \textit{quark propagator in the axial gauge} at a given order in $\alpha_s$~\cite{Bosch:2004th, Becher:2006qw, Bauer:2001yt}. The jet function for the decay of $B$-mesons into light hadronic final states is known both at NLO~\cite{Bosch:2004th} and at NNLO~\cite{Becher:2006qw}.
\item[\textit{iii})] Finally, the \textit{shape function} $S$ is a non-perturbative object describing the internal soft dynamics of the $B$-meson~\cite{Bigi:1993ex,Neubert:1993ch,Neubert:1993um}. Since this function cannot be computed from first principles, it needs to be extracted from the experimental measurement of suitable channels. The leading order shape function $S$ can be estimated from the inclusive radiative decay $B\to X_s\gamma$. Non-perturbative effects at leading power need to be modelled using the most agnostic approach possible so that the uncertainty attached to the lack of knowledge on that object can be quantified and incorporated into the analysis~\cite{Lange:2005yw}.
\end{itemize}
In this work, however, we are only interested in the partonic process underlying the semileptonic $B$-meson decay, namely the bottom quark decay $b\to X_u\ell\bar{\nu}$. Also for the partonic process, HQET+SCET can be employed to prove that the decay width factorises as shown in eq.~\eqref{eq:erSCETFactorisation}, with the difference that in the partonic case, all the functions that appear in the factorization formula can be calculated perturbatively. In particular, in this case, the shape function $S$ is referred to as the soft function and it was evaluated perturbatively up to NNLO in \cite{Becher:2005pd}. In addition, in the partonic decay, the hard, hard-collinear, and soft scales are $m_b$, $\sqrt{m_b m_X}$, and $m_X$, respectively.

It is possible to rewrite eq.~(\ref{eq:erSCETFactorisationWi}) in terms of the $W_i$ functions defined in eq.~(\ref{eq:Wmunu-decomp}). Here we provide the explicit expressions of these functions at NLO as obtained in the EFT. As expected, these expressions reproduce exactly the singular structure of the results in eq.~\eqref{NLO}. We find
\begin{align}\label{eq:winlo}
\widetilde{W}^{(1)}_1 =&  -\bigl(1 - \hat{q}^2\bigr) \left[\frac{\ln(\mxs)}{\mxs}\right]_+\!\! -\bigl(1 -\hat{q}^2\bigr) \Biggl(\frac{7}{4} - 2\ln(1-\hat{q}^2)\Biggr) \left[\frac{1}{\mxs}\right]_+ \!\!
+ \bigl(1-\hat{q}^2\bigr)\, {\cal S}_1 \,\delta(\mxs)
 \, , \nonumber\\
\widetilde{W}^{(1)}_2=& -4  \left[\frac{\ln(\mxs)}{\mxs}\right]_+ - \Biggl(7 - 8\ln(1-\hat{q}^2)\Biggr) \left[\frac{1}{\mxs}\right]_+ + 4\,{\cal S}_2  \,\delta(\mxs) 
\, ,\nonumber\\
\widetilde{W}^{(1)}_3=& -2\left[\frac{\ln(\mxs)}{\mxs}\right]_+ - \Biggl(\frac{7}{2} - 4\ln(1-\hat{q}^2)\Biggr)\left[\frac{1}{\mxs}\right]_+ + 2\,{\cal S}_1\, \delta(\mxs)  
\, ,
\end{align}
where the tilde indicates that only the singular terms of the form factors $W_i$ are included in the r.h.s.\ of the equations and ${\cal S}_i$ are defined in eq.~(\ref{eq:Si}). Notice that the coefficients above are independent of the renormalization scale $\mu$. At NLO the scale dependence is of higher order and only appears in the running of $\alpha_s$. 
NNLO results are too lengthy to be explicitly written here, but we provide them in an ancillary file of the arXiv submission.

The scale dependence of the $W^{(2)}_i$ can be reconstructed from Renormalization Group (RG) equations by using the fact that the hadronic tensor is overall independent of the scale. In addition, we know that the individual $W_i$ are independent of the renormalization scale $\mu$
\begin{align} \label{eq:REGWi}
\frac{d}{d \ln \mu} W_i(\mu) = 0\, .    
\end{align}
From eqs.~\eqref{eq:REGWi} and \eqref{eq:Wexp} it follows that
\begin{align}\label{eq:rgscaledep}
\frac{d}{d \ln \mu} W^{(1)}_i(\mu) = 0\, ,\qquad   \frac{d}{d \ln \mu} W^{(2)}_i(\mu) = \frac{\beta_0}{2}\, W^{(1)}_i \, ,
\end{align}
where $\beta_0=\big(11/3\, C_A - 4/3 \, T_f n_l\big)$ with $C_A=3$, $T_f = 1/2$ and $n_l$ is the number of light flavors.
The RG flow in the second equation of~\eqref{eq:rgscaledep} can be solved and the complete scale dependence of $W^{(2)}_i(\mu)$ can be recovered from the relation
\begin{align}
 W^{(2)}_i(\mu) =  \frac{\beta_0}{2}\, W^{(1)}_i \ln\frac{\mu}{m_b}   +  W^{(2)}_i(m_b)\, ,
\end{align}
provided that $W^{(2)}_i(m_b)$ is known.

In this work we employ the NNLO expression of the form factors in the EFT, $\widetilde{W}^{(2)}_i$  in combination with the BLM contribution to the form factors and the parton level Monte Carlo discussed in Section~\ref{sec:slicing} in order to obtain expressions for the full NNLO form factors, $W^{(2)}_i$, which include the full analytic dependence on the plus distributions and the Dirac delta function  and the analytic dependence on all terms proportional to $\beta_0$,  as well as a numerical fit that describes the contribution of radiation that is neither soft nor collinear.

\section{Slicing Method for NNLO Real Radiation}
\label{sec:slicing}

In the following, we discuss a numerical implementation of NNLO QCD corrections to the partonic process $b\to X_u\ell\bar{\nu}$. Together with the singular results obtained via EFT methods discussed in Section~\ref{sec:SCET} and the NNLO BLM corrections \cite{Gambino:2006wk}, this provides the basis for the fitting procedure which allows us to determine  the structure functions $W^{(2)}_i$. A numerical calculation of the NNLO QCD corrections to the $b$-quark decay in Fermi theory can be obtained by mapping the semileptonic decay of a (massive) $b$ quark onto the process $t\to bW$ and taking the limit $m_b\to 0$ in the top-decay calculation. Indeed, a public parton-level Monte Carlo code that evaluates the NNLO QCD corrections to the decay $t\to bW$ with a massless bottom quark and an on-shell $W$ boson has been available for quite some time~\cite{Gao:2012ja}. This code relies on the analytic calculation of the two-loop corrections to the decay process \cite{Bonciani:2008wf, Asatrian:2008uk, Beneke:2008ei, Bell:2008ws}, and on the numerical evaluation of the one-loop diagrams with an additional gluon in the final state and the tree-level diagrams with two additional gluons in the final state. 

In order to extract numerical fits for the two-loop form factors in eq.~(\ref{eq:Wexp}), in this work we modified the code of \cite{Gao:2012ja} so that the customized version can evaluate the $b\to u\ell\bar{\nu}$ single and double differential decay widths up to NNLO together with the total width. The original code for the evaluation of the top-quark decay was modified in several non trivial ways:

\begin{enumerate}
    \item  An additional integration over the leptonic invariant mass $q^2$ and appropriate factors had to be added in order to obtain the $b$-decay width, since the code described in \cite{Gao:2012ja} evaluated the top decay for an on-shell $W$-boson, while we need to evaluate the $b\to X_u\ell\bar{\nu}$ decay for arbitrary momentum transfer $q^2$.
    
    \item The two-loop corrections to the heavy-to-light current -- essentially the hard function in eq.~(\ref{eq:hardfunc}) -- were evaluated in the code of \cite{Gao:2012ja} at the fixed momentum value $q^2 = M_W^2$. However, in the context of this work, $q^2$ is an event-dependent variable. It was therefore necessary to rewrite the two-loop virtual corrections as a function of $q^2$.

    \item Throughout the original version of the code presented in \cite{Gao:2012ja}, no distinction was made between $q^2$ and $M_W^2$. It was therefore necessary to revisit the code to distinguish the instances in which factors of $M_W^2$ originated from the on-shell condition $q^2 = M_W^2$ employed in the study of the top decay, from instances in which factors of $M_W^2$ were related to the Fermi theory coupling constant.
\end{enumerate}

\noindent In addition, one must observe that the role played by the neutrino in the decay of the top quark corresponds to the role played by the charged lepton in the decay of the bottom quark. Hence, this had to be accounted for in adapting the top-decay code to our purposes.

With this customized and adapted code in hand, one can obtain the differential decay width with 
respect to two variables, for example $\hat{q}_0$ and $\hat{q}^2$ in eq.~(\ref{eq:variables}).
These are indeed the variables on which the three $W_i$ form factors that we need to extract depend 
on. In addition, from the 
code, one can obtain (differential) lepton-energy moments of arbitrary order.

The original top-quark decay code \cite{Gao:2012ja} is based on a slicing scheme which employs $\hat{m}^2_X$ as the resolution variable. In line with the recent developments of non-local subtraction methods  discussed in \cite{Alioli:2025hpa}, we also implemented a non-local subtraction in the code, as an additional feature along with the standard slicing scheme for NNLO calculations. The slicing calculation is carried out by splitting the differential decay rate
\begin{align}
d \Gamma &= d \Gamma^{(0)} +\frac{\alpha_s}{\pi} \, d\Gamma^{(1)} + \bigg(\frac{\alpha_s}{\pi}\bigg)^2 \, d\Gamma^{(2)}  +\ldots \, ,\nonumber \\
&= d \Gamma^{(0)} + d\Gamma^{\delta \mathrm{NLO}} \,\, +  d\Gamma^{\delta \mathrm{NNLO}}  + \ldots \, ,
\end{align}
into two contributions, one where $\mxs$ is below  an infrared (IR) cutoff parameter $\delta\ll 1$, and one where $\mxs$ is larger than $\delta$. Up to subleading power corrections in $\delta$, it is possible to write the NNLO correction to the differential decay rate as,
\begin{align}\label{eq:slic}
    \frac{d\Gamma^{\delta \mathrm{NNLO}}}{d \hat{q}^2 \, d \hat{E}_\ell} &= \frac{d\Gamma^{\mathrm{Sing}} (\delta)}{d \hat{q}^2 \, d \hat{E}_\ell} +\int^{(\hat{m}^{2}_X)^{\mathrm{max}}}_{\delta} d\hat{m}^2_X  \, \frac{d \Gamma_{b\to u\ell\bar{\nu} +\mathrm{jet}}^{\delta\mathrm{NLO}}}{d \hat{q}^2 d \hat{E}_\ell \, d \hat{m}^2_X} + \ldots \, ,
\end{align}
where the dots indicate the missing power corrections in $\delta$. The upper integration limit for $\hat{m}^2_X$ in terms of fixed values of $\hat{E}_\ell$ and $\hat{q}^2$ can be obtained from the phase-space region described by eqs.~\eqref{eq:physrange} and \eqref{eq:elhatregion} and it is given by
\begin{align}
    (\hat{m}^{2}_X)^{\mathrm{max}} = 1-2 \hat{E}_\ell +\hat{q}^2\bigg(1-\frac{1}{2 \hat{E}_\ell}\bigg)\, .
\end{align}
The contribution above the IR cutoff $\delta$ in eq.~\eqref{eq:slic} is provided by the NLO calculation of the $b\to u\ell\bar{\nu}+\mathrm{jet}$ process
\begin{align}\label{eq:nloplusjet}
\frac{d \Gamma_{b\to u\ell\bar{\nu} +\mathrm{jet}}^{\delta \mathrm{NLO}}}{d \hat{q}^2 d \hat{E}_\ell \, d \hat{m}^2_X} &= \int d \Phi_4 \, V(\Phi_4) \,\delta(\hat{q}^{\prime\, 2}(\Phi_4) - \hat{q}^2 )\, \delta(\hat{E}^{\prime}_\ell(\Phi_4) - \hat{E}_\ell)\,\delta(\hat{m}^{\prime\, 2}_X(\Phi_4)-\hat{m}^2_X) \, \nonumber \\
&+ \int  d \Phi_5 \, R(\Phi_5) \,\delta(\hat{q}^{\prime\, 2}(\Phi_5) - \hat{q}^2 )\, \delta(\hat{E}^{\prime}_\ell(\Phi_5) - \hat{E}_\ell)\,\delta(\hat{m}^{\prime\, 2}_X(\Phi_5)-\hat{m}^2_X) \, ,
\end{align}
where $\Phi_4$ and $\Phi_5$ represent the four- and five-particle phase spaces. The quantity $V(\Phi_4)$ represents the one-loop corrections to the $b\to u\ell\bar{\nu} +\mathrm{jet}+X$ decay process, and the quantity $R(\Phi_5)$ represents the real emission corrections to the $b\to u\ell\bar{\nu} +\mathrm{jet}+X$ process. The triple differential distribution is obtained through the Dirac delta functions in eq.~\eqref{eq:nloplusjet} which evaluate the kinematic functions $\hat{q}^{\prime\, 2}$, $\hat{E}^\prime_\ell$ and $\hat{m}^{\prime\, 2}_X$ on the phase spaces $\Phi_4$ and $\Phi_5$ to the values $\hat{q}^2$, $\hat{E}_\ell$ and $\mxs$. The quantity
\begin{align}\label{eq:cumulant}
\frac{d\Gamma^{\mathrm{Sing}}(\delta)}{d \hat{q}^2 \, d \hat{E}_\ell}  \equiv\int_0^{\delta} d \hat{m}^2_X  \, \frac{d \Gamma}{d \hat{q}^2 d \hat{E}_\ell \, d \hat{m}^2_X} \bigg|_{\mathrm{Sing}\, \mathcal{O}(\alpha^2_s)}\, ,
\end{align}
represents the leading power singular cumulative decay rate at $\mathcal{O}(\alpha_s^2)$ differential in $\hat{q}^2$ and $\hat{E}_\ell$ which can be obtained at $\mathcal{O}(\alpha_s^2)$ from eq.~(\ref{eq:erSCETFactorisation}) by integrating the singular limit of the decay width over $\hat{m}^2_X$. The singular cumulant is different from zero in the range of values for $\hat{q}^2$ and $\hat{E}_\ell$ that correspond to Born phase-space configurations ($m^2_X = 0$).

It is important to notice that the r.h.s. of eq.~(\ref{eq:slic}) is valid up to power corrections in $\delta$ since the cumulative cross section at $\mathcal{O}(\alpha_s^2)$ below the IR cutoff is approximated by its singular expression. Hence, the IR parameter $\delta$ must be chosen carefully, in order to avoid distortions in the distributions due to the missing power corrections. The NLO calculation, above the $\delta$ cutoff, for the $b\to u\ell\bar{\nu}+\mathrm{jet}$ process is evaluated numerically through the dipole subtraction method \cite{Catani:1996vz,Melnikov:2011qx}. The implementation of the $\hat{E}_\ell$ moments in the NLO $b\to u\ell\bar{\nu}+\mathrm{jet}$ calculation is obtained by multiplying the differential decay rate by the appropriate power of $\hat{E}_\ell$.

It is also possible to rewrite eq.~(\ref{eq:slic}) as a nonlocal subtraction formula, similarly to the approach taken in \cite{Alioli:2025hpa}, by introducing a second cutoff parameter $\hat{m}^{2\, \mathrm{cut}}_X\ge \delta$
\begin{align}\label{eq:nonlocal}
\frac{d\Gamma^{\delta \mathrm{NNLO}}}{d \hat{q}^2 \, d \hat{E}_\ell} &= \frac{d\Gamma^{\mathrm{Sing}}(\hat{m}^{2\, \mathrm{cut}}_X)}{d \hat{q}^2 \, d \hat{E}_\ell}  \, \nonumber\\
&+ \int^{(\hat{m}^{2}_X)^{\mathrm{max}}}_{\delta} d\hat{m}^2_X \, \bigg[\frac{d \Gamma_{b\to u\ell\bar{\nu} +\mathrm{jet}}^{\delta \mathrm{NLO}}}{d \hat{q}^2 d \hat{E}_\ell \, d \hat{m}^2_X}  -  \frac{d \Gamma}{d \hat{q}^2 d \hat{E}_\ell \, d \hat{m}^2_X} \bigg|_{\mathrm{Sing}\, \mathcal{O}(\alpha^2_s)} \!\! \theta(\hat{m}^{2\, \mathrm{cut}}_X - \hat{m}^2_X)\bigg] + \ldots\, ,    
\end{align}
where $\delta$ now plays the role of a technical cutoff parameter and could be, in principle, sent to zero, provided that the virtual-real and double-real amplitudes are numerically stable in the infrared regions of QCD. This is typically never the case, hence the IR cutoff $\delta$, in a nonlocal subtraction, is usually set to a small but finite value. The second cutoff parameter $\hat{m}^{2\, \mathrm{cut}}_X$ can be freely chosen in the range $\delta\le \hat{m}^{2\, \mathrm{cut}}_X \le (\hat{m}^2_X)^{\mathrm{max}}$ and, in particular, if one sets $\hat{m}^{2\, \mathrm{cut}}_X=\delta$ in eq.~(\ref{eq:nonlocal}) one recovers the slicing formula in eq.~(\ref{eq:slic}). Both the singular cumulant in the first line of eq.~(\ref{eq:nonlocal}) and the singular subtraction term in the line below are nonvanishing for values of $\hat{q}^2$ and $\hat{E}_\ell$ that correspond to Born kinematics configurations. This condition achieves the proper cancellation between the cumulant and the singular subtraction term in eq.~(\ref{eq:nonlocal}) in the region $\delta < \hat{m}^2_X < \hat{m}^{2\, \mathrm{cut}}_X$. In addition, the numerical calculations needed in the following sections are carried out by setting $\delta=5\times 10^{-5}$ and $\hat{m}_X^{2\, \mathrm{cut}}=10^{-2}$.

At a more practical level, to extract the form factors means fitting functions ($W_1, W_2$ and $W_3$ mentioned above) of two variables ($\hat{q}_0$, $\hat{q}^2$) to the numerical results of the code for the decay width and the lepton energy moments. This is complicated by the fact that the functions which are to be fitted are singular in some corners of the phase space. However, using the EFT results, it is possible to evaluate analytically the singular terms which can then be accounted for exactly, as discussed in Section~\ref{sec:SCET}. 

In conclusion, by running the parton-level Monte Carlo on a computer cluster, it was possible to obtain 
a grid of numerical values for the differential decay width of the semileptonic decay of the bottom quark (eq.~\eqref{eq:double}) at NNLO. In addition, the numerical code also allowed us to obtain the first two lepton energy moments (see eqs.~\eqref{eq:double1} and \eqref{eq:double2}) at NNLO accuracy. Those numerical distributions were employed to extract the numerical fits for the NNLO structure functions, as discussed in Section~\ref{sec:form-factors}.

\section{Structure Functions Extraction\label{sec:form-factors}}

In this section we describe the fitting strategy employed to extract the non-singular contributions to the hadronic structure functions from numerical results for several differential distributions and integrated observables. Our fits rely primarily on the double-differential distribution in $\hat{m}_X^2$ and $\hat{q}^2$ and on its first two $\hat{E}_\ell$ moments, which we compute with the numerical approach of Section~\ref{sec:slicing}. As additional constraints, we include the total decay rate~\cite{vanRitbergen:1999gs} and the first two total lepton-energy moments~\cite{Pak:2008qt}, which are available analytically. Some fits, including our final results, also incorporate the $\hat{q}^2$ spectrum, constructed by integrating the analytic expression found in~\cite{Chen:2022wit} over suitable $\hat{q}^2$ bins.

We use the following notation for the observables included in the fit:
\begin{equation}\label{eq:obs_fits}
\left\langle\dfrac{d^2 M_{m}}{d\hat{m}_X^2\, d\hat{q}^2}\right\rangle_{[\hat{m}_{X}^2]_i,\hat{q}^2_j} \, ,\quad
\left\langle\dfrac{d\Gamma}{d\hat{q}^2}\right\rangle_{[\hat{q}^2]_i}\, ,\quad
M_m\, ,
\end{equation}
where $m=0,1,2$ labels the $\hat{E}_\ell$ moments. The notation $M_m$ provides a compact way of denoting the decay width and its first two lepton-energy moments,
\begin{equation}\label{eq:moms_fits_notation}
M_0 = \Gamma\, ,\qquad
M_1 = \langle E_\ell \rangle\, ,\qquad
M_2 = \langle E_\ell^2 \rangle\, .
\end{equation}
Unless stated otherwise, the symbols in eqs.~\eqref{eq:obs_fits} and \eqref{eq:moms_fits_notation} refer to the contribution at a fixed perturbative order (NLO or NNLO). To keep the notation light, in this section we suppress an explicit perturbative-order label in $M_m$, since the procedure is applied independently at each order.

In eq.~\eqref{eq:obs_fits}, $[\hat{m}_{X}^2]_i=\left[\hat{m}_{X,i}^2, \hat{m}_{X,i+1}^2\right]$ and $[\hat{q}^2]_i=\left[\hat{q}_i^2, \hat{q}_{i+1}^2\right]$, with $\hat{m}_{X,i}^2 < \hat{m}_{X,i+1}^2$ and $\hat{q}_i^2 < \hat{q}_{i+1}^2$, denote bins in $\hat{m}_X^2$ and $\hat{q}^2$, respectively. The subscript $\hat{q}^2_j$ in the double-differential distributions indicates that they are evaluated at a fixed value $\hat{q}^2=\hat{q}^2_j$.

The double-differential distributions are evaluated only for non-zero values of $\hat{m}_X^2$, since the computational scheme in eq.~\eqref{eq:nonlocal} selects exclusively the NLO contribution to the $b\to u\ell\bar{\nu} + \mathrm{jet}$ process. The first bin, which includes $\hat{m}_X^2 = 0$, is excluded from the fit, as it is dominated by the known singular contribution.

Given a generic function $\mathcal{F}(\hat{E}_\ell,\hat{q}_0,\hat{q}^2)$, we define the bin averages appearing in eq.~\eqref{eq:obs_fits} as
\begin{align}
&\left\langle \mathcal{F}(\hat{E}_\ell, \hat{q}_0, \hat{q}^2)\right\rangle_{[\hat{m}_{X}^2]_i,\hat{q}^2_j}
= \int_{\hat{m}_{X,i}^2}^{\hat{m}_{X,i+1}^2} d\hat{m}_X^2
\int_{\hat{E}_\ell^\mathrm{min}(\hat{m}_{X}^2,\hat{q}^2_j)}^{\hat{E}_\ell^\mathrm{max}(\hat{m}_{X}^2,\hat{q}^2_j)} d\hat{E}_{\ell}\;
\tilde{\mathcal{F}}(\hat{E}_\ell, \hat{m}_X^2, \hat{q}^2)\Big|_{\hat{q}^2=\hat{q}^2_j}\, ,\nonumber\\[3pt]
&\left\langle \mathcal{F}(\hat{E}_\ell, \hat{q}_0, \hat{q}^2)\right\rangle_{[\hat{q}^2]_i}
= \int_{\hat{q}_i^2}^{\hat{q}_{i+1}^2} d\hat{q}^2
\int_{\hat{m}_X^{2,\mathrm{min}}(\hat{q}^2)}^{\hat{m}_X^{2,\mathrm{max}}(\hat{q}^2)} d\hat{m}_X^2
\int_{\hat{E}_\ell^\mathrm{min}(\hat{m}_{X}^2,\hat{q}^2)}^{\hat{E}_\ell^\mathrm{max}(\hat{m}_{X}^2,\hat{q}^2)} d\hat{E}_{\ell}\;
\tilde{\mathcal{F}}(\hat{E}_\ell, \hat{m}_X^2, \hat{q}^2)\, ,\nonumber\\[3pt]
&\left\langle \mathcal{F}(\hat{E}_\ell, \hat{q}_0, \hat{q}^2)\right\rangle
= \int_{\hat{q}^{2,\mathrm{min}}}^{\hat{q}^{2,\mathrm{max}}} d\hat{q}^2
\int_{\hat{m}_X^{2,\mathrm{min}}(\hat{q}^2)}^{\hat{m}_X^{2,\mathrm{max}}(\hat{q}^2)} d\hat{m}_X^2
\int_{\hat{E}_\ell^\mathrm{min}(\hat{m}_{X}^2,\hat{q}^2)}^{\hat{E}_\ell^\mathrm{max}(\hat{m}_{X}^2,\hat{q}^2)} d\hat{E}_{\ell}\;
\tilde{\mathcal{F}}(\hat{E}_\ell, \hat{m}_X^2, \hat{q}^2)\, ,
\end{align}
where $\tilde{\mathcal{F}}(\hat{E}_\ell,\hat{m}_X^2,\hat{q}^2)=\mathcal{F}\!\left(\hat{E}_\ell,\hat{q}_0(\hat{m}_X^2,\hat{q}^2),\hat{q}^2\right)$ and $\hat{q}_0$ is the normalised dilepton total energy,
\begin{equation}
\hat{q}_0 = \dfrac{1 + \hat{q}^2 - \hat{m}_X^2}{2}\, .
\end{equation}
The quantities $\hat{E}_\ell^\mathrm{min/max}$, $\hat{m}_X^{2,\mathrm{min/max}}$ and $\hat{q}^{2,\mathrm{min/max}}$ denote the kinematic boundaries (see eqs.~\eqref{eq:physrange} and~\eqref{eq:elhatregion}).

It is now useful to return to the structure of the $b \to u\ell\bar{\nu}$ form factors $W_i$. At each perturbative order, we schematically decompose the structure functions as
\begin{equation}\label{eq:sing_reg_W_i_decomp}
W_i^{(n)}(\hat{q}_0, \hat{q}^2) = \widetilde{W}_i^{(n)}(\hat{q}_0,\hat{q}^2) + R_i^{(n)}(\hat{q}_0, \hat{q}^2)\, ,
\end{equation}
where $\widetilde{W}_i^{(n)}$ collects the singular contributions, which can be computed within an EFT (see Section~\ref{sec:SCET}), and $R_i^{(n)}$ denotes the remaining regular terms. At NLO, both $\widetilde{W}_i^{(1)}$ and the remainder $R_i^{(1)}$ are known analytically (the latter corresponds to the $\mathcal{R}_i^{(1)}$ functions in eq.~\eqref{NLO}). 

At NNLO, the decomposition is analogous, but additional analytic information is available: the BLM contributions to the structure functions are known analytically. We therefore further  split the NNLO form factors as follows
\begin{equation}
W_i^{(2)}(\hat{q}_0, \hat{q}^2) =
\widetilde{W}^{(2)}_i(\hat{q}_0, \hat{q}^2) +
R^{(2,\beta_0)}_i(\hat{q}_0, \hat{q}^2) +
R_i^{(2)}(\hat{q}_0, \hat{q}^2)\,,
\label{eq:splitWs}
\end{equation}
where $R^{(2,\beta_0)}_i$ denotes the regular part of $W_i^{(2)}$ proportional to $\beta_0$. The first two terms on the r.h.s. of eq.~\eqref{eq:splitWs} are known analytically, while the non-BLM regular terms, $R_i^{(2)}$ are not; consequently they are determined via the fitting procedure described below. The fits for  $R_i^{(2)}$ are the main new results of this work.

The binned distributions and moments used as numerical inputs contain both singular and regular contributions. In practice, we isolate the part that is sensitive to the unknown regular pieces by subtracting, at the observable level, the contributions induced by the analytically known singular terms (and, at NNLO, by the regular BLM pieces). The fit is then performed on the resulting ``regular-only'' data, which depend linearly on the unknown functions $R_i^{(n)}$. 

We model the \emph{regular} part of the $b\to X_u\ell\bar{\nu}$ structure functions as a linear combination of basis functions,
\begin{equation}\label{eq:R_i_basis of functions}
R_i^{(n)}(\hat{q}_0, \hat{q}^2) = \sum_{j=1}^N \alpha^{(n)}_{ij} f^{(n)}_j(\hat{q}_0, \hat{q}^2)\, ,
\end{equation}
where $N$ is the number of model functions $f^{(n)}_j(\hat{q}_0,\hat{q}^2)$. To keep the notation lighter in the rest of this section, we suppress the superscript $(n)$ in $\alpha_{ij}$ and $f_j$; this does not introduce any ambiguity since the fit is performed order by order.

The contribution of the $W_i$ regular terms to the observables in eq.~\eqref{eq:obs_fits} can then be written as
\begin{align}
\left\langle\dfrac{d^2 M_{m}}{d\hat{m}_X^2\, d\hat{q}^2}\right\rangle_{[\hat{m}_{X}^2]_i,\hat{q}^2_j}^{reg}
&= \dfrac{G_F^2 m_b^5|V_{ub}|^2}{16\pi^3}
\sum_{k=1}^3 \sum_{\ell=1}^N
\alpha_{k\ell}\Big\langle\beta_{k\ell}^{m}\Big\rangle_{[\hat{m}_{X}^2]_i,\hat{q}^2_j}\, ,\nonumber\\[2pt]
\left\langle\dfrac{d\Gamma}{d\hat{q}^2}\right\rangle_{[\hat{q}^2]_i}^{reg}
&= \dfrac{G_F^2 m_b^5|V_{ub}|^2}{16\pi^3}
\sum_{k=1}^3 \sum_{\ell=1}^N
\alpha_{k\ell}\Big\langle\gamma_{k\ell}\Big\rangle_{[\hat{q}^2]_i}\, ,\nonumber\\[2pt]
M_{m}^{reg}
&= \dfrac{G_F^2 m_b^5|V_{ub}|^2}{16\pi^3}
\sum_{k=1}^3 \sum_{\ell=1}^N
\alpha_{k\ell}\Big\langle\delta_{k\ell}^{m}\Big\rangle\, ,
\end{align}
in terms of \emph{preprocessing integrals} that depend only on the choice of basis functions and on the kinematic binning. Here $k=1,2,3$ labels the three structure functions $W_{1,2,3}$ that contribute for massless leptons (cf. eq.~\eqref{eq:triple}). The preprocessing integrals are defined as
\begin{align}\label{eq:preprocessing}
&\Big\langle\beta_{k\ell}^{m}\Big\rangle_{[\hat{m}_{X}^2]_i,\hat{q}^2_j} = \int_{\hat{m}_{X,i}^2}^{\hat{m}_{X,i+1}^2} d\hat{m}_X^2\int_{\hat{E}_\ell^\mathrm{min}}^{\hat{E}_\ell^\mathrm{max}} d\hat{E}_{\ell}\,\hat{E}_{\ell}^m\,\tilde{g}_k(\hat{E}_\ell,\hat{m}_X^2,\hat{q}^2)\,\tilde{f}_\ell(\hat{m}_X^2,\hat{q}^2)\Bigg|_{\hat{q}^2=\hat{q}^2_j}\, ,\nonumber \\
&\Big\langle\gamma_{k\ell}\Big\rangle_{[\hat{q}^2]_i} = \int_{\hat{q}^2_i}^{\hat{q}^2_{i+1}} d\hat{q}^2\int_{\hat{m}_X^{2,\mathrm{min}}}^{\hat{m}_X^{2,\mathrm{max}}} d\hat{m}_X^2 \int_{\hat{E}_\ell^\mathrm{min}}^{\hat{E}_\ell^\mathrm{max}} d\hat{E}_{\ell}\, \tilde{g}_k(\hat{E}_\ell,\hat{m}_X^2,\hat{q}^2)\, \tilde{f}_\ell(\hat{m}_X^2,\hat{q}^2)\, ,\nonumber \\
&\Big\langle\delta_{k\ell}^{m}\Big\rangle = \int_{\hat{q}^{2,\mathrm{min}}}^{\hat{q}^{2,\mathrm{max}}} d\hat{q}^2 \int_{\hat{m}_X^{2,\mathrm{min}}}^{\hat{m}_X^{2,\mathrm{max}}} d\hat{m}_X^2 \int_{\hat{E}_\ell^\mathrm{min}}^{\hat{E}_\ell^\mathrm{max}} d\hat{E}_{\ell}\,\hat{E}_{\ell}^m\, \tilde{g}_k(\hat{E}_\ell,\hat{m}_X^2,\hat{q}^2)\, \tilde{f}_\ell(\hat{m}_X^2,\hat{q}^2)\,.
\end{align}
In these expressions we defined $\tilde{f}_\ell(\hat{m}_X^2,\hat{q}^2) = f_\ell\left(\hat{q}_0(\hat{m}_X^2,\hat{q}^2),\hat{q}^2\right)$, and $\tilde{g}_k(\hat{E}_\ell,\hat{m}_X^2,\hat{q}^2) = g_k\left(\hat{E}_\ell, \hat{q}_0(\hat{m}_X^2,\hat{q}^2),\hat{q}^2\right)$ with
\begin{equation}
g_k(\hat{E}_\ell, \hat{q}_0, \hat{q}^2) = \left\{
\begin{array}{ccc}
&\hat{q}^2\, , & k=1\\
&-\left(2 \hat{E}_{\ell}^2-2\hat{E}_{\ell}\hat{q}_0 +\frac{\hat{q}^2}{2}\right)\, , & k=2\\
&\hat{q}^2 (2\hat{E}_\ell-\hat{q}_0)\, , & k=3
\end{array}
\right.
\end{equation}
Once these preprocessing integrals are computed, predictions for the fitted observables follow immediately for any set of parameters $\alpha_{ij}$, and the best-fit values are obtained by comparison with the numerical inputs.

We evaluate the preprocessing integrals using a multidimensional adaptive integration routine implemented in \texttt{JAX}~\cite{jax2018github,Frostig:2018}. The implementation is optimized for vectorized evaluations and provides error estimates associated with both the quadrature procedure and the finite number of integrand evaluations.

A practical advantage of keeping track of the numerical uncertainties from the preprocessing step is that we can assign \emph{model errors} by standard error propagation. For instance,
\begin{align}
\sigma^2\Bigg[\left\langle\dfrac{d^2 M_{m}}{d\hat{m}_X^2\, d\hat{q}^2}\right\rangle_{[\hat{m}_{X}^2]_i,\hat{q}^2_j}\Bigg]
=
\left(\dfrac{G_F^2 m_b^5|V_{ub}|^2}{16\pi^3}\right)^2
\sum_{k=1}^3 \sum_{\ell=1}^{m}\alpha_{k\ell}^2
\,\sigma^2\Bigg[\Big\langle\beta_{k\ell}^{m}\Big\rangle_{[\hat{m}_{X}^2]_i,\hat{q}^2_j}\Bigg] \, .
\end{align}

\noindent The dataset used for the NLO and NNLO extractions discussed in the following two sections consists of:
\begin{itemize}
\item[\textit{i})] the double-differential distribution $\dfrac{d^2 M_{m}}{d\hat{m}_X^2\, d\hat{q}^2}$ and the first two lepton-energy moments, evaluated in 200 $\hat{m}_X^2$ bins and at 25 $\hat{q}^2$ points;
\item[\textit{ii})] the single-differential distribution $\dfrac{d\Gamma}{d\hat{q}^2}$, evaluated in 100 $\hat{q}^2$ bins;
\item[\textit{iii})] the total rate and the first two total lepton-energy moments.
\end{itemize}
Given this dataset and the corresponding model predictions, we construct a $\chi^2$ statistic that incorporates both data and integration uncertainties:
\begin{equation}\label{eq:chi2_data_errs}
\chi^2(\theta_j)=\sum_{i=1}^{n_\mathrm{obs}}\dfrac{(y_i - \lambda_i(\theta_i))^2}{\sigma_{y_i}^2 + \sigma_{\lambda_i}^2(\theta_j)}\, ,
\end{equation}
where $y_i$ denotes the numerical inputs, $\sigma_{y_i}$ their associated uncertainties, $\lambda_i$ the model predictions, $\theta_j$ the set of fit parameters (in our case the $\alpha_{ij}$), and $\sigma_{\lambda_i}$ the model uncertainties induced by the numerical evaluation of the preprocessing integrals.

\section{NLO Structure Functions}
\label{sec:NLO}

At NLO, exact analytic results for the double differential distribution have been known for a long time (see e.g.~\cite{DeFazio:1999ptt}), and analytic expressions for the structure functions $W^{(1)}_i$ can be found in Section~\ref{sec:InclusiveNLO}. Here we study the NLO corrections to the form factors $W_i$ with the method which we also apply to NNLO. A direct comparison of the results obtained with our semi-analytic method to the exact analytic results available at NLO allows us to validate the method and to see which kind of performance one can reasonably expect at NNLO, where analytic results for the double differential distributions in $\hat{q}^2$ and $\mxs$ are not available.

To start, for a set of fixed values of $\hat{q}^2$, we ran the parton-level Monte Carlo code to obtain NLO distributions binned in $\mxs$. These data, the total NLO $b$-quark decay width, and the first two total lepton energy moments at NLO, $\langle E^n_l \rangle$ with $n=1,2$ \cite{Pak:2008qt}, are then employed to obtain fits for the $W_i^{(1)}$ ($i=1,2,3$) structure functions. To be specific, only $R^{(1)}_i$,  the part of the  $W_i^{(1)}$ structure functions which cannot be calculated with EFT methods, as described in Section~\ref{sec:SCET}, is fitted. The part of the $W^{(1)}_i$ functions that can be obtained by means of EFT methods is instead accounted for analytically in all of the results in this work.

The monomials employed to build the fitting functions are chosen by requiring that they respect the analytic structure of the known analytic result at NLO. We chose functions that have a cut for $-\infty < \hat{q}^2 < 0$ or for $-\sqrt{\hat{q}^2} < \hat{q}_0 < \sqrt{\hat{q}^2}$. We also added monomials corresponding to subleading power corrections (i.e. terms proportinal to powers of $\hat{q}^2$ and $\mxs$), and we limited ourselves to transcendental functions of maximum weight one (i.e. we consider at most single logarithms). Specifically, the model chosen to fit the NLO structure functions $W_i^{(1)}$ is
\begin{equation}
R_i^{(1)} (\mxs,\hat{q}^2) = \sum_{j=1}^{14} \alpha_{ij}^{(1)} f^{(1)}_{j}(\mxs,\hat{q}^2)  \, ,
\end{equation}
where $\alpha^{(1)}_{ij}$ are the fitting constants and the monomials $f^{(1)}_j$ are part of the set
\begin{align}
f^{(1)}_j \in& \Bigl\{1, \ln\mxs, \mxs, \mxs \ln\mxs, \hat{q}^2, \hat{q}^2 \ln\mxs, \hat{q}^4, \hat{q}^4 \ln\mxs, \nonumber\\&
\ln(1 -\hat{q}^2), {\mathcal I}_1, \hat{q}^2 \ln(1 -\hat{q}^2), \hat{q}^2 {\mathcal I}_1, \hat{q}^4 \ln(1 -\hat{q}^2), \hat{q}^2 {\mathcal I}_1 \Bigr\}\,. \label{eq:modelNLO}
\end{align}
The quantity ${\mathcal I}_1$ was already introduced in eq.~(\ref{eq:I_1}). 

\begin{figure}[tp]
\begin{center}
\includegraphics[width=0.68\textwidth]{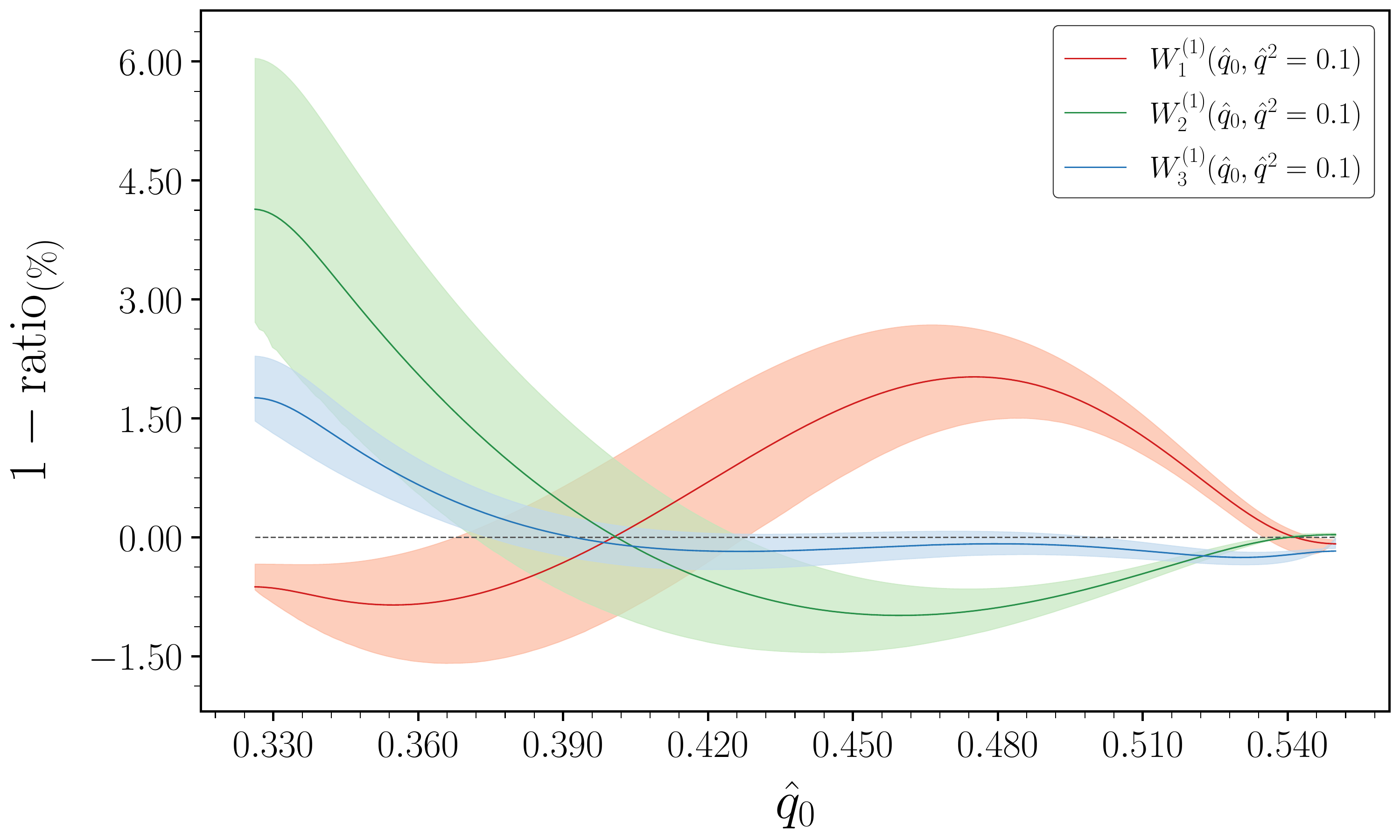}\\
\includegraphics[width=0.68\textwidth]{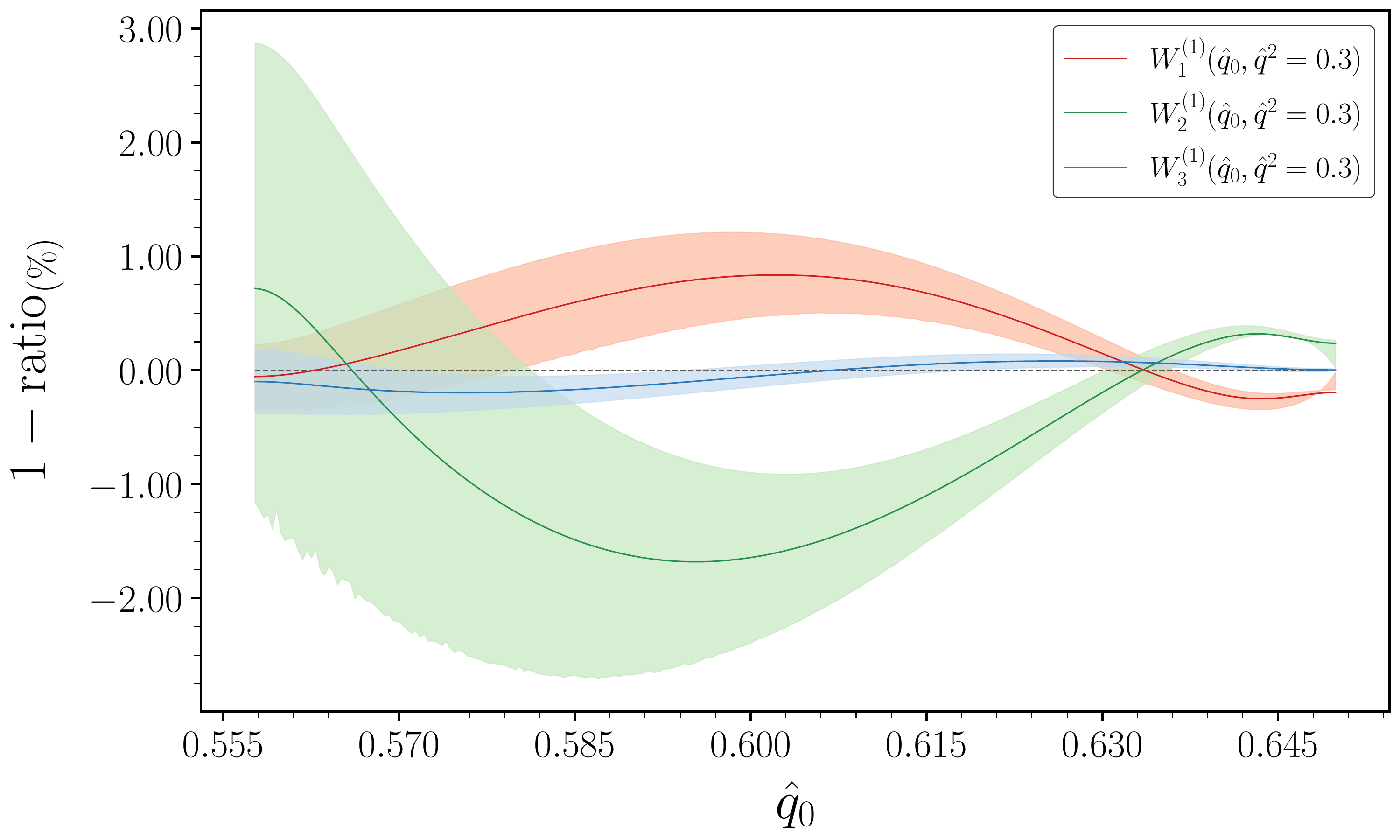}\\
\includegraphics[width=0.68\textwidth]{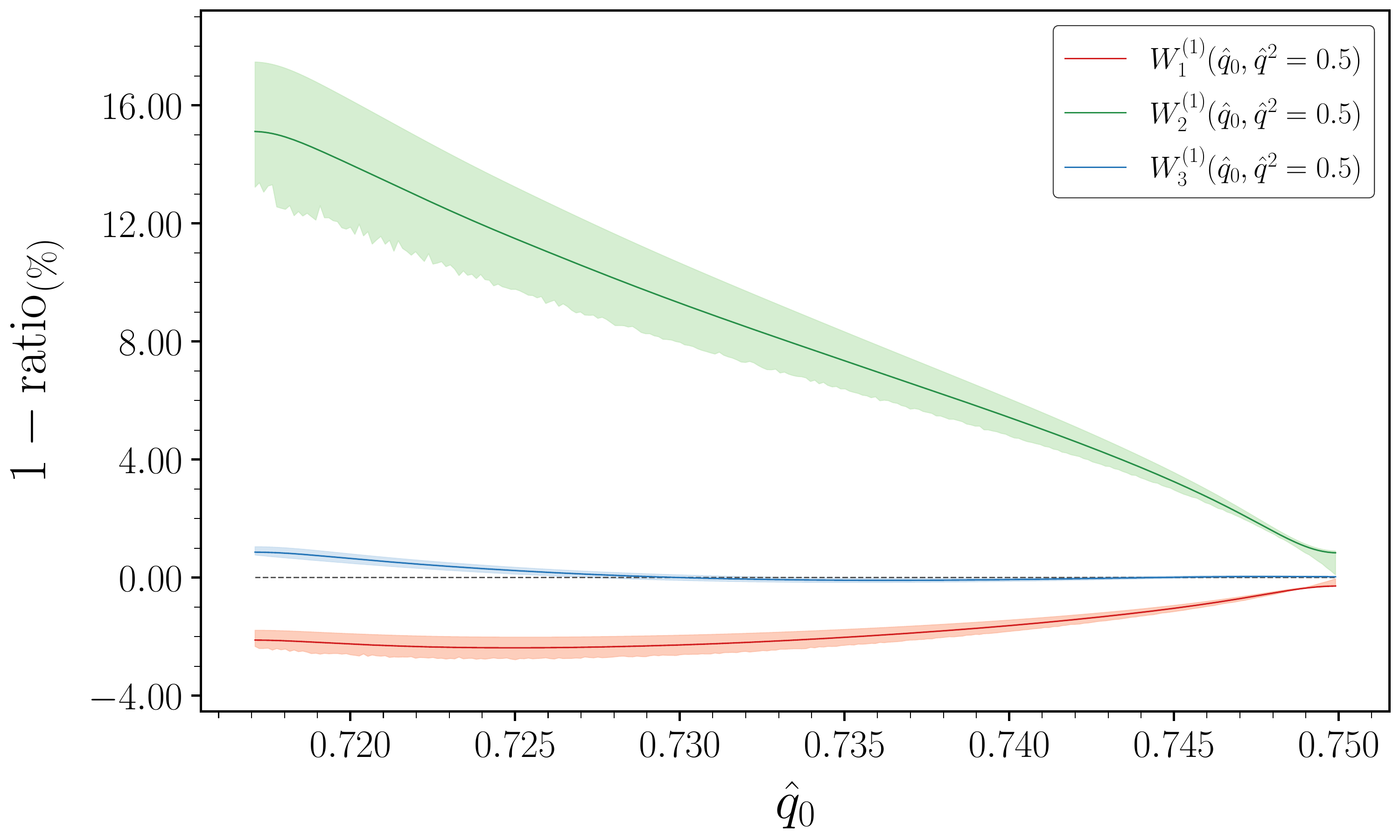}
\caption{$\mathcal{O}(\alpha_s)$ contributions to the hadronic structure functions $W_i$ as functions of kinematic variables, for three representative values of the lepton invariant mass, $\hat{q}^2=0.1,0.3,0.5$. The bands represent the 68\% CL ranges from the fit, which includes the $\hat{q}^2$ differential distribution. The vertical axes show $1-\mathrm{ratio}$, where one takes the ratio of the fitted $W_i^{(1)}$ (numerator) to the analytic result (denominator).}
\label{fig:Wi1_q2h}
\end{center}
\end{figure}

After the values of the fitting constants $\alpha^{(1)}_{ij}$ have been determined through the fitting procedure described in Section~\ref{sec:form-factors}, one can compare the fitted form factors $W_i^{(1)}$ to their analytic expression and/or use the fitted structure functions to obtain one-parameter differential partonic distributions that can then be compared with known results.

Figure~\ref{fig:Wi1_q2h} shows the behavior of the NLO form factors $W_i^{(1)}$ as functions of $\hat{q}_0$, at fixed values $\hat q^2 = 0.1$ (upper panel), $\hat q^2 = 0.3$ (central panel) and $\hat q^2 = 0.5$ (lower panel). Each panel contains three curves corresponding to $W^{(1)}_1, W^{(1)}_2, W^{(1)}_3$. 

By observing Figure~\ref{fig:Wi1_q2h},  one can conclude that the analytic expressions of the $W^{(1)}_i$ form factors and the corresponding fits generally agree well, within about 2\% percent. However, larger deviations are observed for $W^{(1)}_2$ at $\hat{q}^2 = 0.1$ and $0.5$, at low values of $\hat{q}_0$. This can be understood as follows: in the double differential distributions employed in the fit, $W_2$ is multiplied by a factor $(\hat{q}_0^2 - \hat{q}^2)= \vec q^{\;2}$, see eqs.~\eqref{eq:double}, \eqref{eq:double1}, and \eqref{eq:double2}, which vanishes at  $\hat{q}_0=\sqrt{\hat{q}^2}$, the minimum allowed value of $\hat{q}_0$. Therefore, the fit cannot determine precisely $W^{(1)}_2$ for $\hat{q}_0\sim\sqrt{\hat{q}^2}$, and this holds for any observable based on the triple differential rate of eq.~(\ref{eq:triple}), as in that region the prefactor of $W_2$ in that equation is quadratic $\vec{q}^{\;2}$. For the very same reasons, however, the uncertainty on $W_2$ in that region has a negligible impact on the calculation of {\it any} physical observable. Moreover, from eq.~\eqref{eq:physrange} we see that, as $\hat{q}^2$ grows, the range of values available for $\hat{q}_0$ shrinks rapidly, which explains why the discrepancy in  $W^{(1)}_2$ is large for most values of $\hat{q}_0$ in the lowest panel of Figure~\ref{fig:Wi1_q2h}. A similar, albeit weaker, suppression around $\hat{q}_0\sim\sqrt{\hat{q}^2}$ occurs for the prefactor $W_3$.

Figure~\ref{fig:Wi1_q2h} also presents the 68\% CL bands obtained in fit. They represent the uncertainty following from the information included in the fit. As expected, the uncertainty on $W_2^{(1)}$ (green band) is largest at low  $\hat{q}_0$. One may wonder why the fit often struggles to reproduce the exact results within the bands: this is due to the limitations  of the model eq.~(\ref{eq:modelNLO}) employed in the fit. By extending the functional basis one can easily improve the agreement between fit and exact results, but we refrain from doing that in order to have a relatively small functional basis, which will be particularly useful in  our NNLO analysis. Therefore the 68\% CL uncertainties provided by the fit do not represent the total uncertainty of our fit results. A more reliable way to assess the overall uncertainty is to compare fit-based predictions of observables with the exact NLO results. We now proceed with this comparison.

\begin{figure}[htp]
\centering
\includegraphics[width=0.71\textwidth]{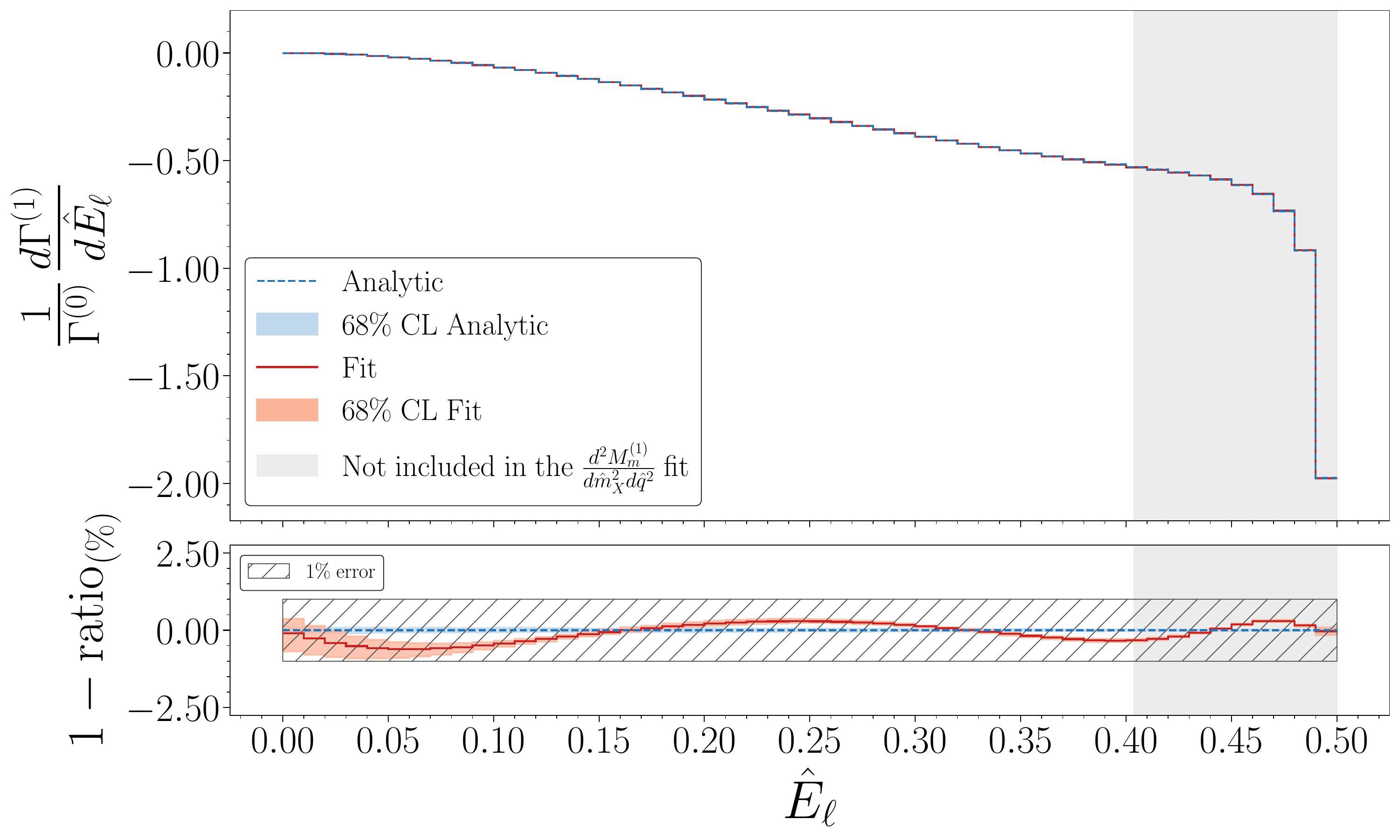}
\\
\includegraphics[width=0.71\textwidth]{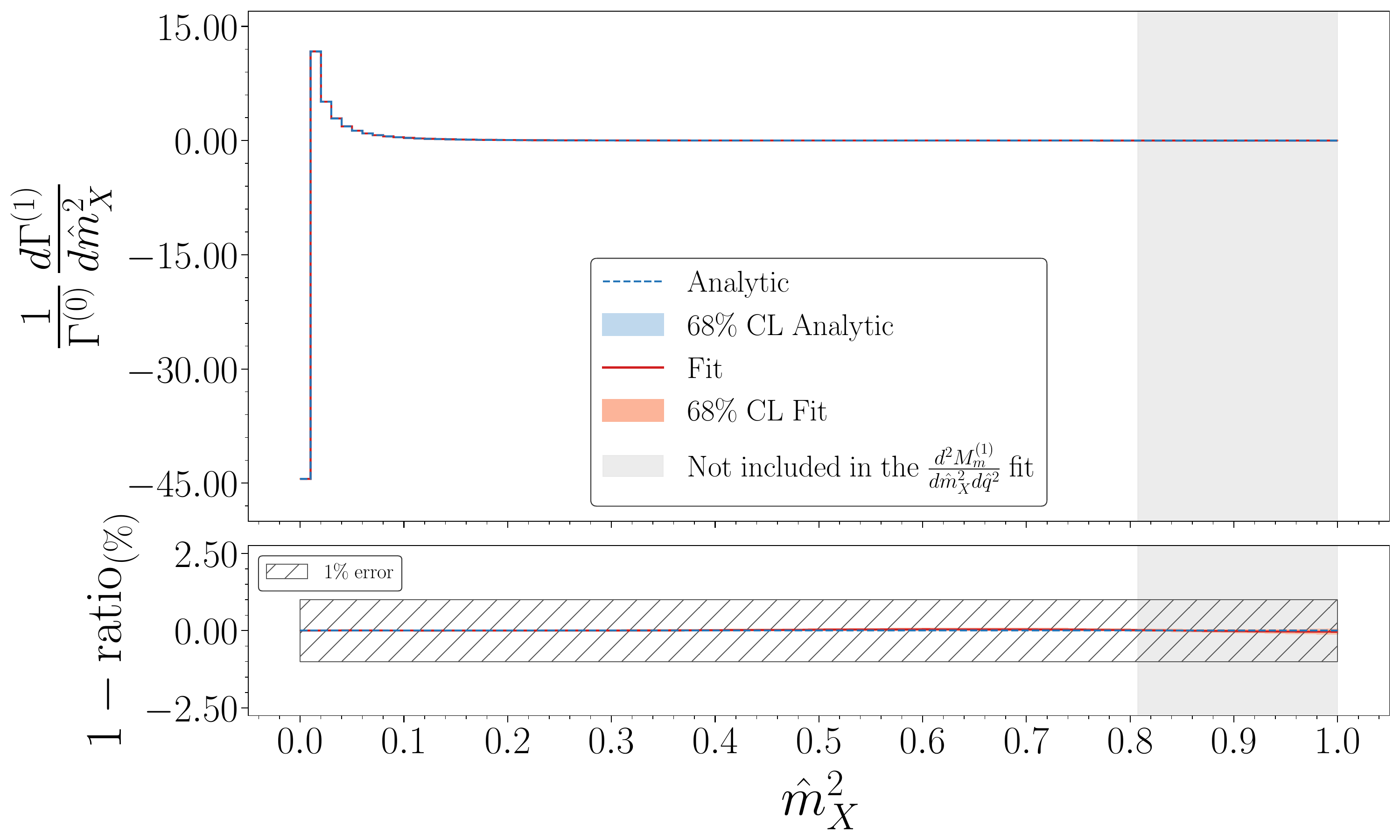}
\\
\includegraphics[width=0.71\textwidth]{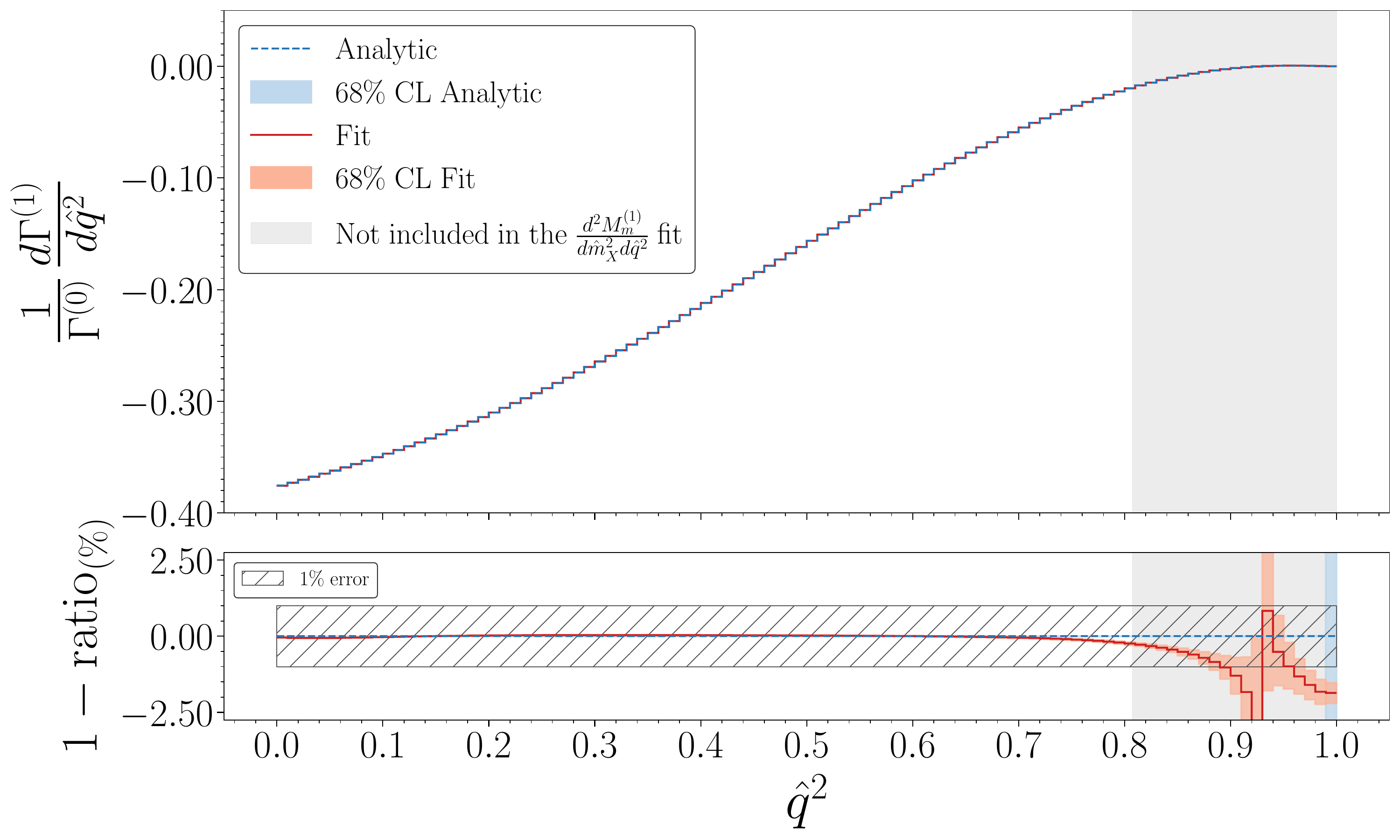}
\caption{NLO differential distributions for the $b \to X_u\ell\bar{\nu}$ decay and comparison with the corresponding results from fitted $W^{(1)}_i$, see text for an explanation. The fits employed in this figure are obtained without including information coming from the analytic $\hat{q}^2$ distribution. }
\label{fig:NLOfits}
\end{figure}

A first example is provided by Figure~\ref{fig:NLOfits}, which compares the NLO corrections to partonic differential distributions obtained from the analytic calculation with the same quantities computed on the basis of the fitted NLO form factors $W^{(1)}_i$. 

\begin{figure}[htp]
\centering
\includegraphics[width=0.71\textwidth]{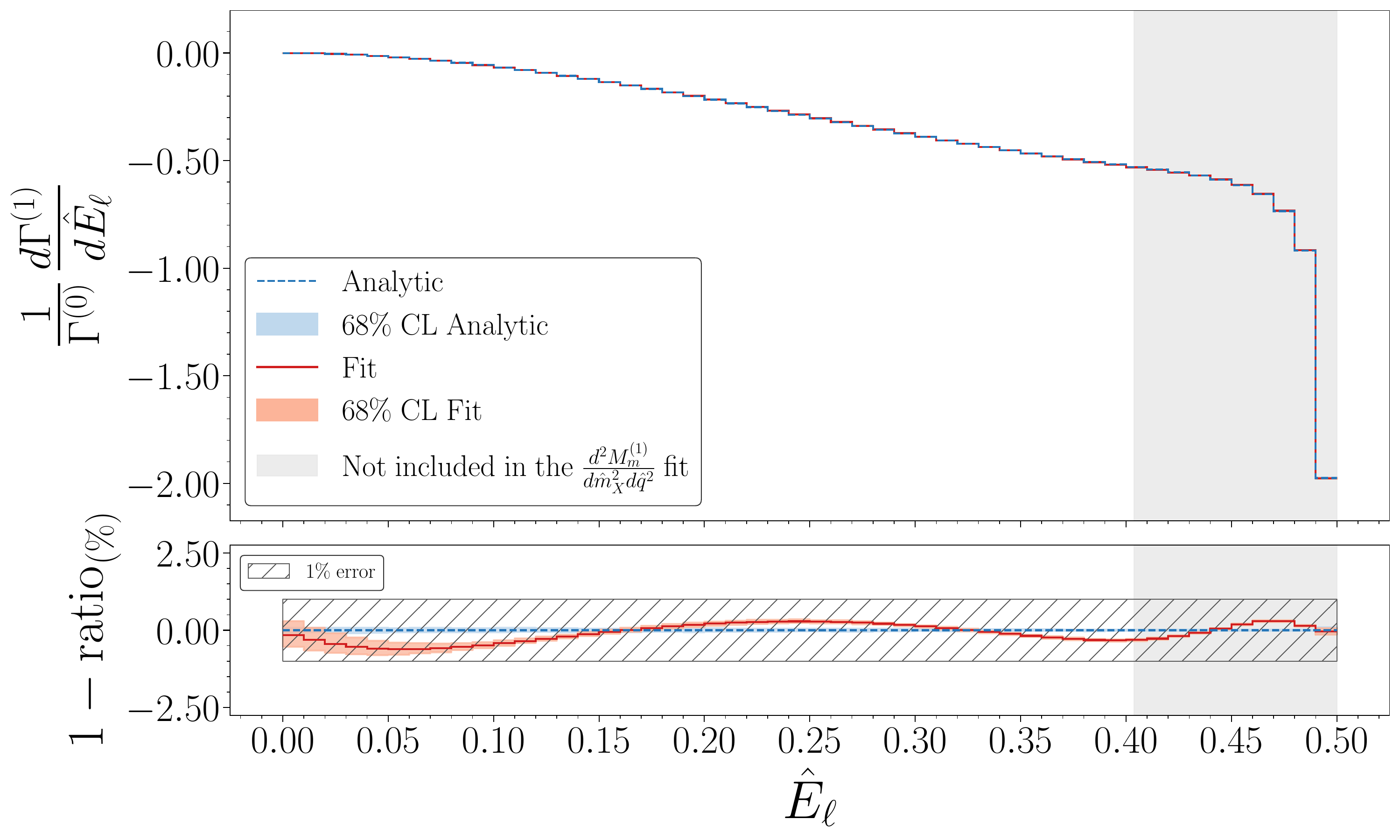}
\\
\includegraphics[width=0.71\textwidth]{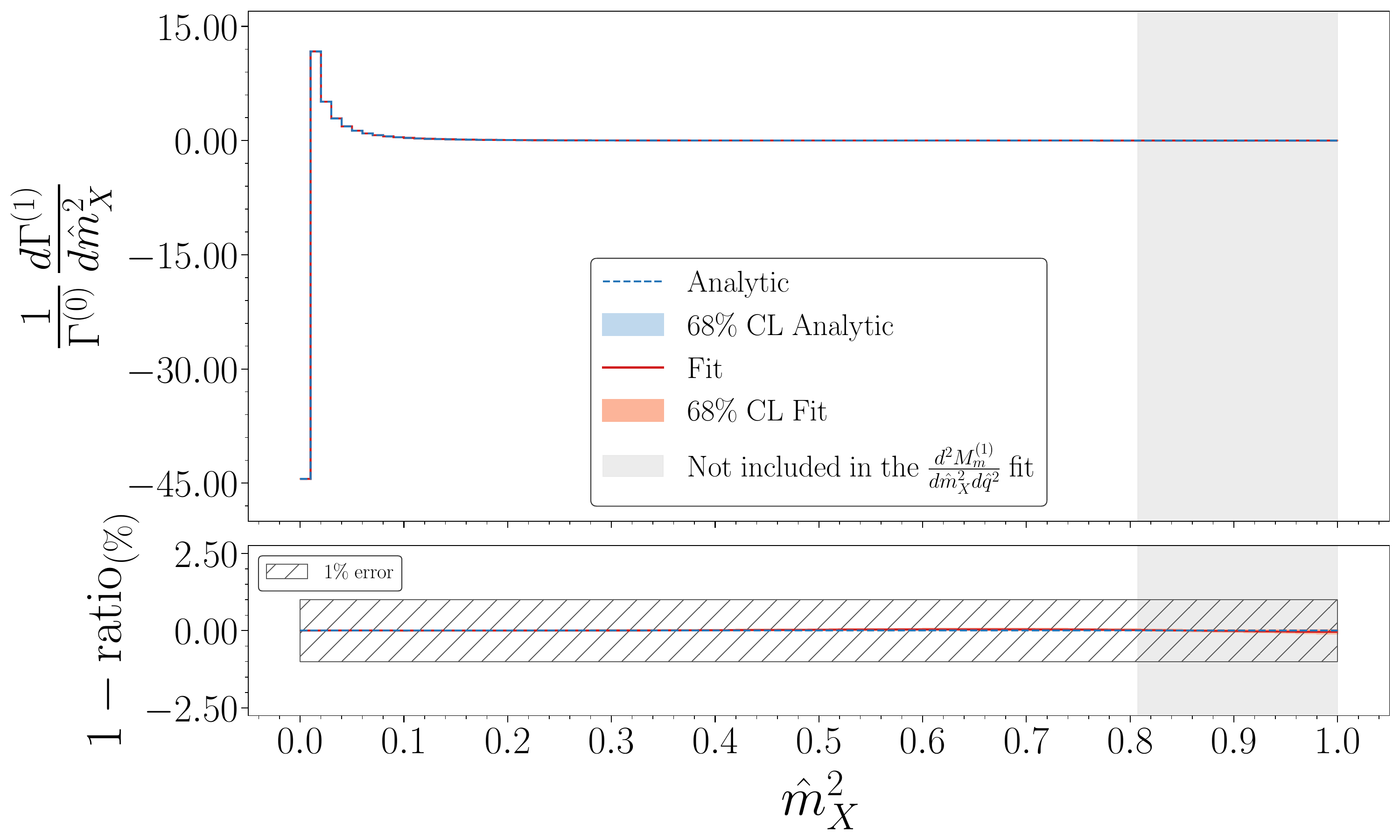}
\\
\includegraphics[width=0.71\textwidth]{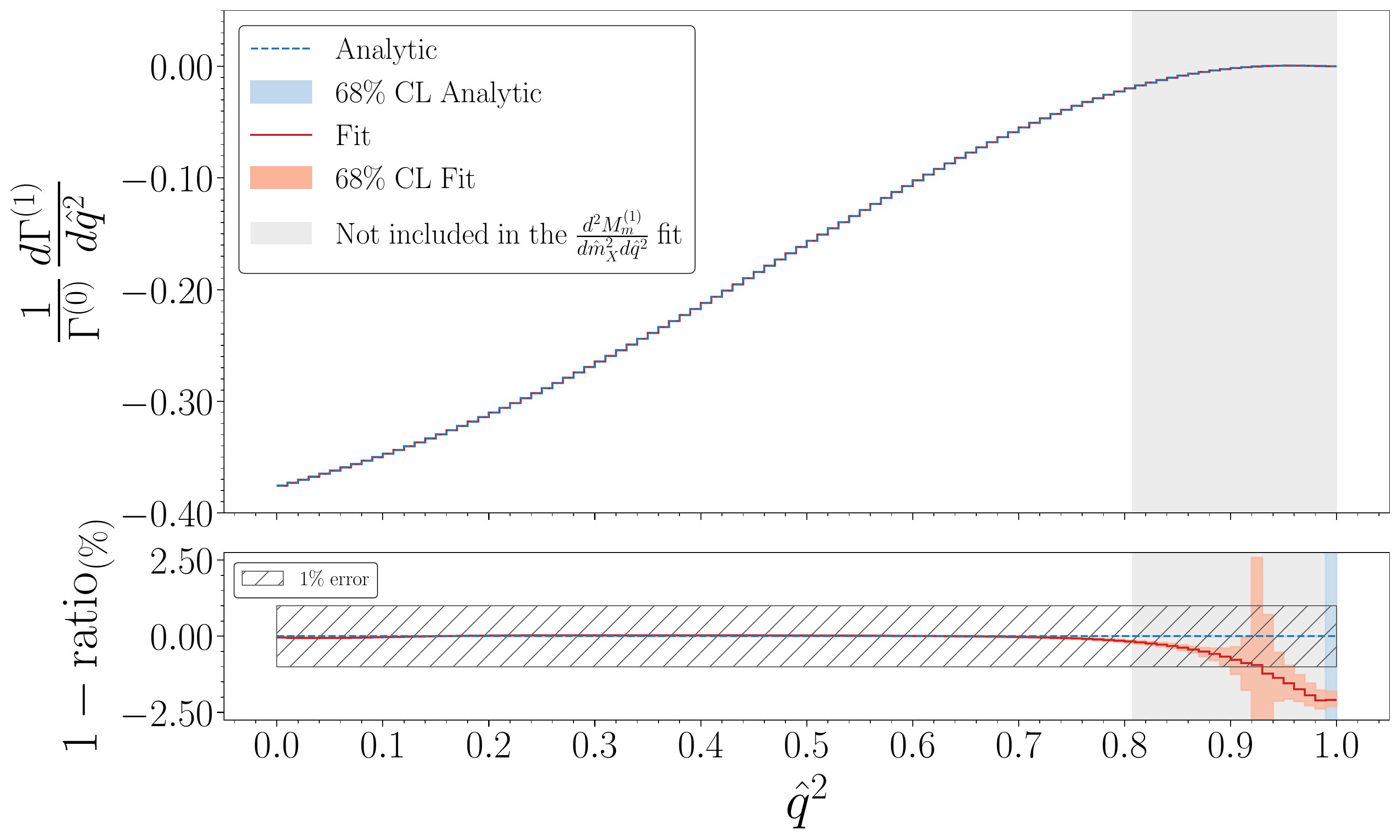}
\caption{NLO differential distributions for the $b \to X_u\ell\bar{\nu}$ decay and comparison with the corresponding results from fitted $W^{(1)}_i$ (see text). The fits employed in this figure include  information coming from the analytic $\hat{q}^2$ distribution. }
\label{fig:NLOfitswithq2h}
\end{figure}

The upper panel in Figure~\ref{fig:NLOfits} shows the NLO correction to the charged lepton energy spectrum, $d\Gamma^{(1)}/d \hat{E}_l $, which corresponds to the order $\alpha_s$ corrections to the triple distribution in eq.~\eqref{eq:triple}, integrated over $\hat{q}^2$ and $\hat{q}_0$. The analytic result is shown by a dashed blue line, while the result based on the fits is shown  with a continuous red line. The uncertainty associated to the determination of the fit parameters is shown by a red band surrounding the red line. The uncertainty affecting the integrated analytic result  is due only to the phase space integration, and it is shown by a (very narrow) blue band. The gray band on the right side of the plot represents a small phase-space region where we did not have Monte Carlo data to constrain the form factors fits. This is due the fact that for high values of $\hat{q}^2$, the range of allowed values for $\mxs$ becomes very small, as it can be seen from eq.~\eqref{eq:physrange}. Consequently in that kinematic region the Monte Carlo provides data on very few bins. The analytic and fitted NLO corrections to the $\hat{E}_l$ spectrum are indistinguishable by eye. The ratio of the fitted result over the analytic result is shown in the inlay below the first panel in Figure~\ref{fig:NLOfits}. There, one can see that the result obtained by the fit and its uncertainty, shown by the red band, is within $1 \%$ of the analytic result in the entire range of $\hat{E}_l$.

The panel in the middle of Figure~\ref{fig:NLOfits} shows the NLO corrections to the differential distribution with respect to $\mxs$. The color and styles are chosen as in the first panel of Figure~\ref{fig:NLOfits}. In this case, the difference between the analytic and fitted result is barely perceptible even when one plots the ratio of the two corrections; the difference remains well below $1 \%$ in the entire physical range.

Finally, the lower panel of Figure~\ref{fig:NLOfits} shows NLO corrections to the $\hat{q}^2$ spectrum. The color code is again the same as in the other panels in Figure~\ref{fig:NLOfits}. The ratio plot in the lowest panel of Figure~\ref{fig:NLOfits} shows that the analytic and fitted corrections differ by less than $1 \%$  with the exception of the region $\hat{q}^2 \in [0.9,1.00]$, where there are no numerical data from the double differential distributions to constrain the fit and the corrections are vanishingly small.

At this stage, it is important to stress that the analytic expression of the $\hat{E}_l$, $\mxs$ and $\hat{q}^2$ spectra were \emph{not} employed to determine the values of the fit parameters employed in Figure~\ref{fig:NLOfits}.

However, since an analytic expression for the $\hat{q}^2$ spectrum is available also at NNLO,  we also produced a NLO fit which includes data from the analytic $\hat{q}^2$ spectrum. Figure~\ref{fig:NLOfitswithq2h} shows the same three differential distributions shown in Figure~\ref{fig:NLOfits} but produced with the fit that includes information from the analytic $\hat{q}^2$ spectrum. One can see that, apart for a marginal improvement of the agreement between fits and analytic results in the high $\hat{q}^2$ region of the $\hat{q}^2$ distribution, the same observations that apply to Figure~\ref{fig:NLOfits} can be made about Figure~\ref{fig:NLOfitswithq2h}. 

We can conclude that the method proposed in this work and the fitting procedure, when applied to the NLO form factors, gives excellent results that reproduce in a very satisfactory way known analytic NLO results. It makes then sense to apply the same procedure to the NNLO corrections to the $W^{(2)}_i$ form factors.

\section{NNLO fits}
\label{sec:NNLO}

We now turn to the main results of this work: the extraction of the structure functions $W_i$ at $\mathcal{O}(\alpha_s^2)$. 

As explained in Section~\ref{sec:SCET}, the NNLO  two-loop corrections are known analytically. In addition, the BLM NNLO corrections are also known analytically. Both these set of corrections are accounted for exactly in the expressions for the form factors $W_i^{(2)}$ that we obtain in this work. Only the NNLO corrections to $W_i^{(2)}$ which are not singular or BLM, $R_i^{(2)}$, are fitted numerically. 

The functional form employed to fit what is left of the form factors was chosen requiring the same analyticity properties required for the NLO fit. However, at NNLO we also include in the functional form  $\ln^2$ and $\ln^3$ functions. Specifically, the ansatz that we use is
\begin{equation}
R_i^{(2)} (\mxs,\hat{q}^2) = \sum_{j=1}^{32} \alpha_{ij}^{(2)} f^{(2)}_{j}(\mxs,\hat{q}^2)\,,
\end{equation}
where
\begin{align}
f_j^{(2)} \in& \Bigl\{1, \ln\mxs, \ln^2\mxs, \ln^3\mxs, \mxs, \mxs \ln\mxs, \mxs \ln^2\mxs, \mxs \ln^3\mxs, \hat{q}^2 \ln\mxs, \nonumber\\
&\hat{q}^2 \ln^2\mxs, \hat{q}^2 \ln^3\mxs, \hat{q}^4 \ln\mxs, \hat{q}^4 \ln^2\mxs, \hat{q}^4 \ln^3\mxs, \ln(1-\hat{q}^2), \ln^2(1-\hat{q}^2), \nonumber\\
& \ln^3(1-\hat{q}^2), \mathcal{I}_1, \mathcal{I}_2, \mathcal{I}_3,\hat{q}^2 \ln(1-\hat{q}^2), \hat{q}^2 \ln^2(1-\hat{q}^2), \hat{q}^2 \ln^3(1-\hat{q}^2), \hat{q}^2 \mathcal{I}_1, \hat{q}^2 \mathcal{I}_2, \hat{q}^2 \mathcal{I}_3,\nonumber \\
& \hat{q}^4 \ln (1-\hat{q}^2), \hat{q}^4 \ln^2(1-\hat{q}^2), \hat{q}^4 \ln^3(1-\hat{q}^2), \hat{q}^4 \mathcal{I}_1, \hat{q}^4 \mathcal{I}_2, \hat{q}^4 \mathcal{I}_3\Bigr\} \, .
\label{eq:NNLOfs}
\end{align}
The numerical coefficients $\alpha_{ij}^{(2)}$ are fixed by the fit. 
In eq.~\eqref{eq:NNLOfs} we introduced ${\mathcal I}_2$ and ${\mathcal I}_3$, two functions related to ${\mathcal I}_1$ through the definition
\begin{align}
{\mathcal I}_n &= \frac{1}{2 \sqrt{\hat{q}_0^2 - \hat{q}^2}} \ln^n \left(\frac{1 - \hat{q}_0 +\sqrt{\hat{q}_0^2 - \hat{q}^2} }{ 1 - \hat{q}_0 - \sqrt{\hat{q}_0^2 - \hat{q}^2}}\right) = \frac{1}{\sqrt{\lambda}} \ln^n \left(\frac{1+ \mxs  -\hat{q}^2 + \sqrt{\lambda} }{1+ \mxs  -\hat{q}^2  - \sqrt{\lambda}  } \right) \, ,
\end{align}
where $n \in \{1,2,3\}$ and $\lambda$ was introduced in eq.~(\ref{eq:lambda}).

The quantities $R^{(2)}_i$ are then combined with the singular and BLM contributions according to eq.~\eqref{eq:splitWs} to obtain semi-analytic expressions for the structure functions $W^{(2)}_i$.

\begin{figure}[htp]
\centering
\includegraphics[width=0.71\textwidth]{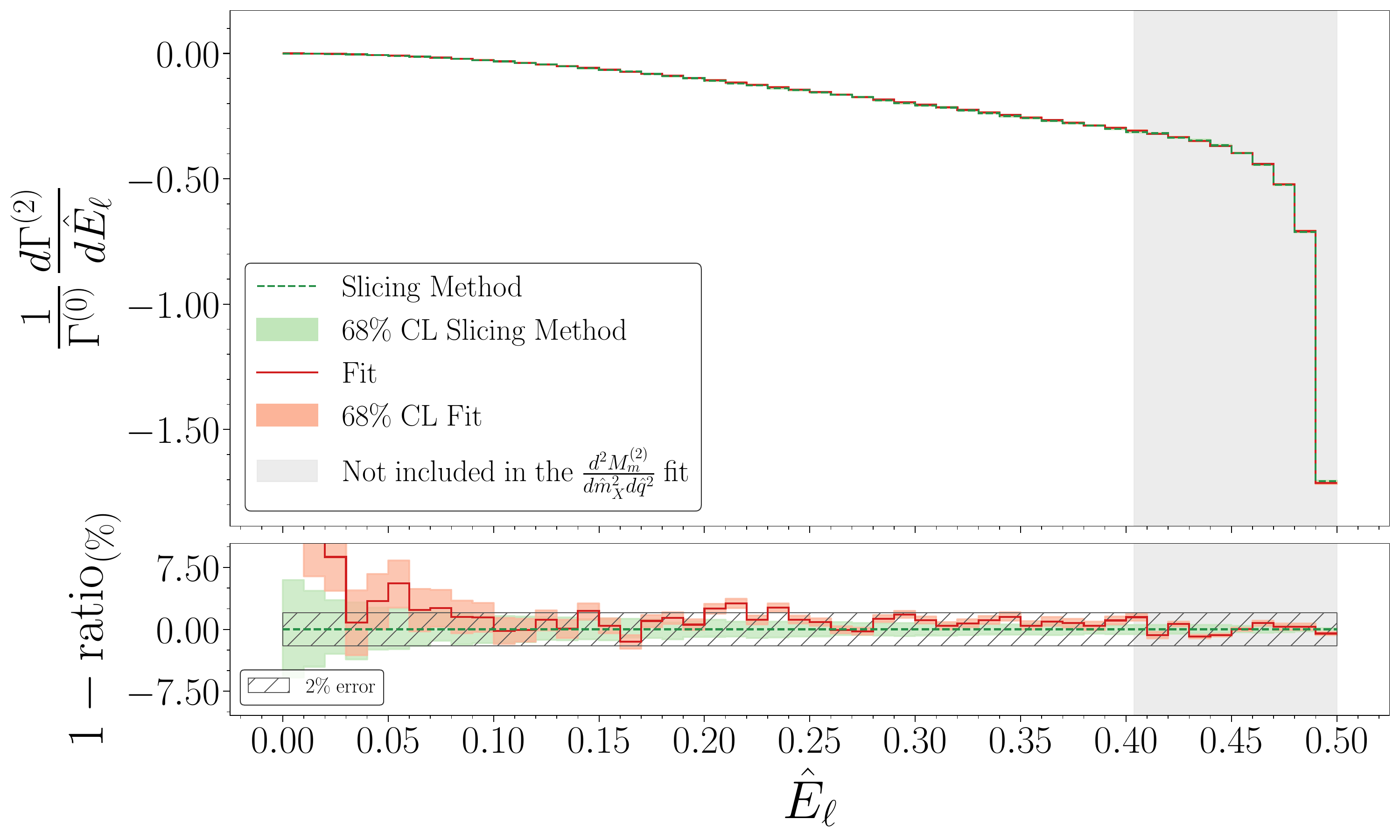}
\\
\includegraphics[width=0.71\textwidth]{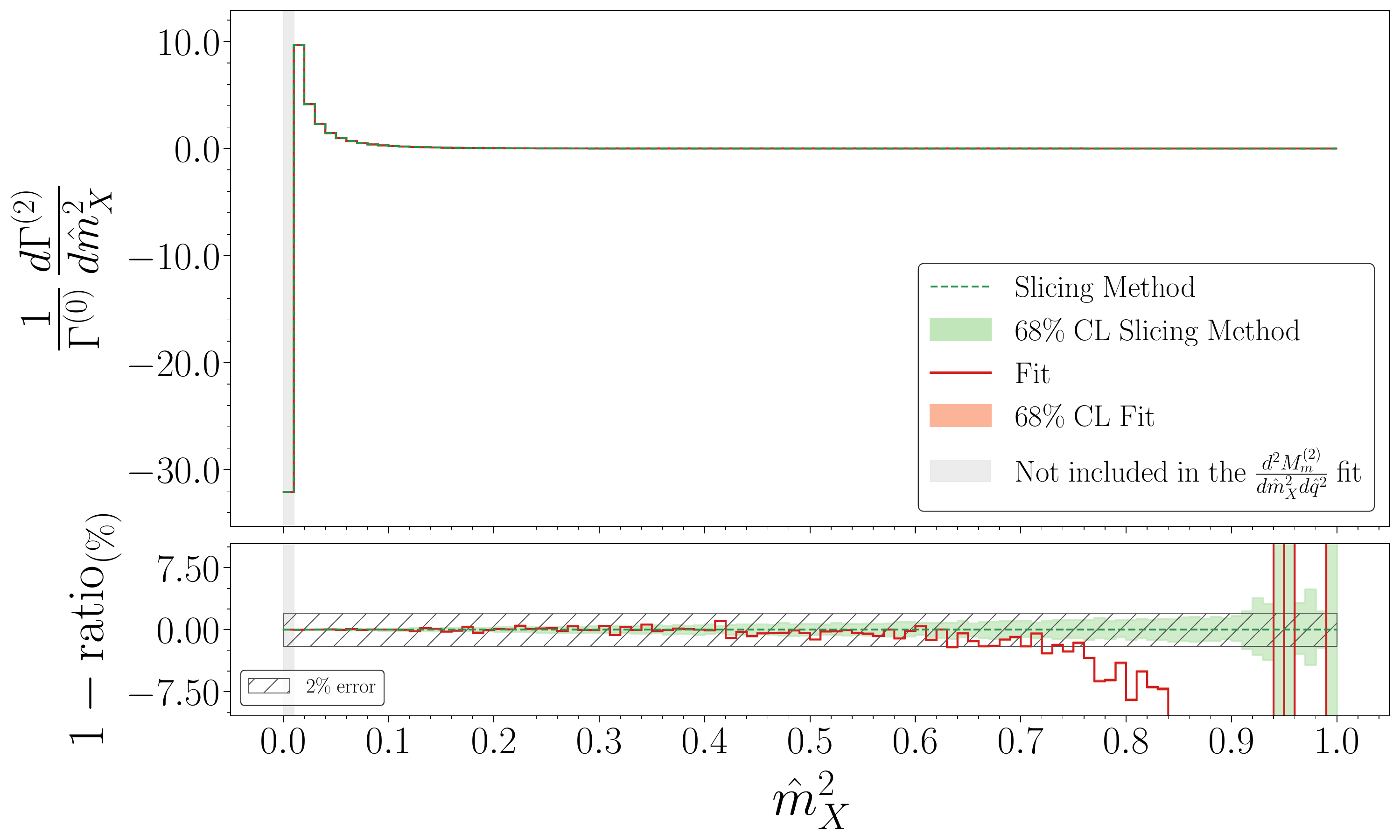}
\\
\includegraphics[width=0.71\textwidth]{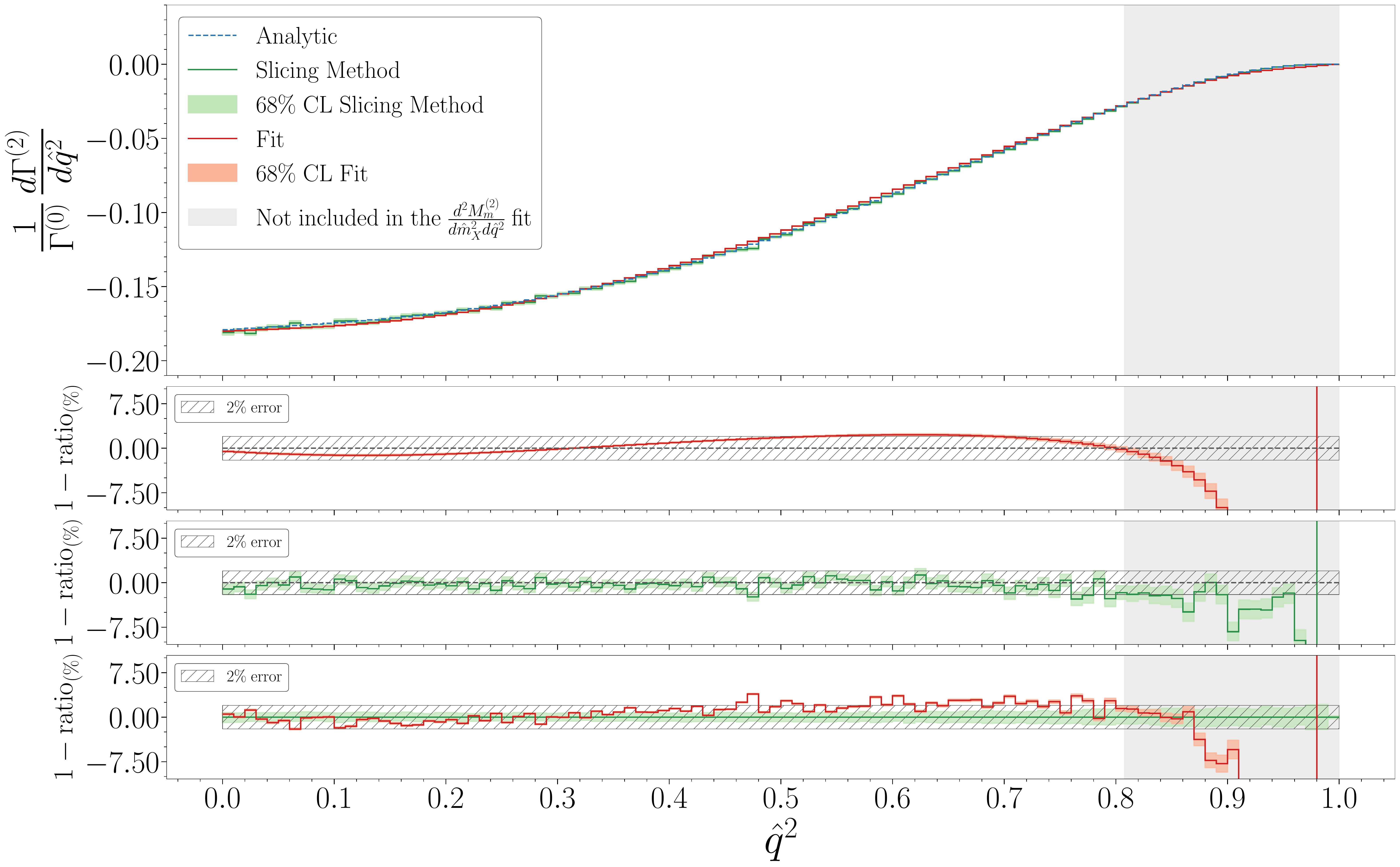}
\caption{NNLO differential decay distributions in $b\to X_u\ell\bar{\nu}$ and comparison with the corresponding results from fitted $W_i^{(2)}$. The fits employed in this Figure are obtained without using information from the analytic results for the $\hat{q}^2$ distribution.}
\label{fig:NNLOfitswithOUTq2h}
\end{figure}

\begin{figure}[htp]
\centering
\includegraphics[width=0.71\textwidth]{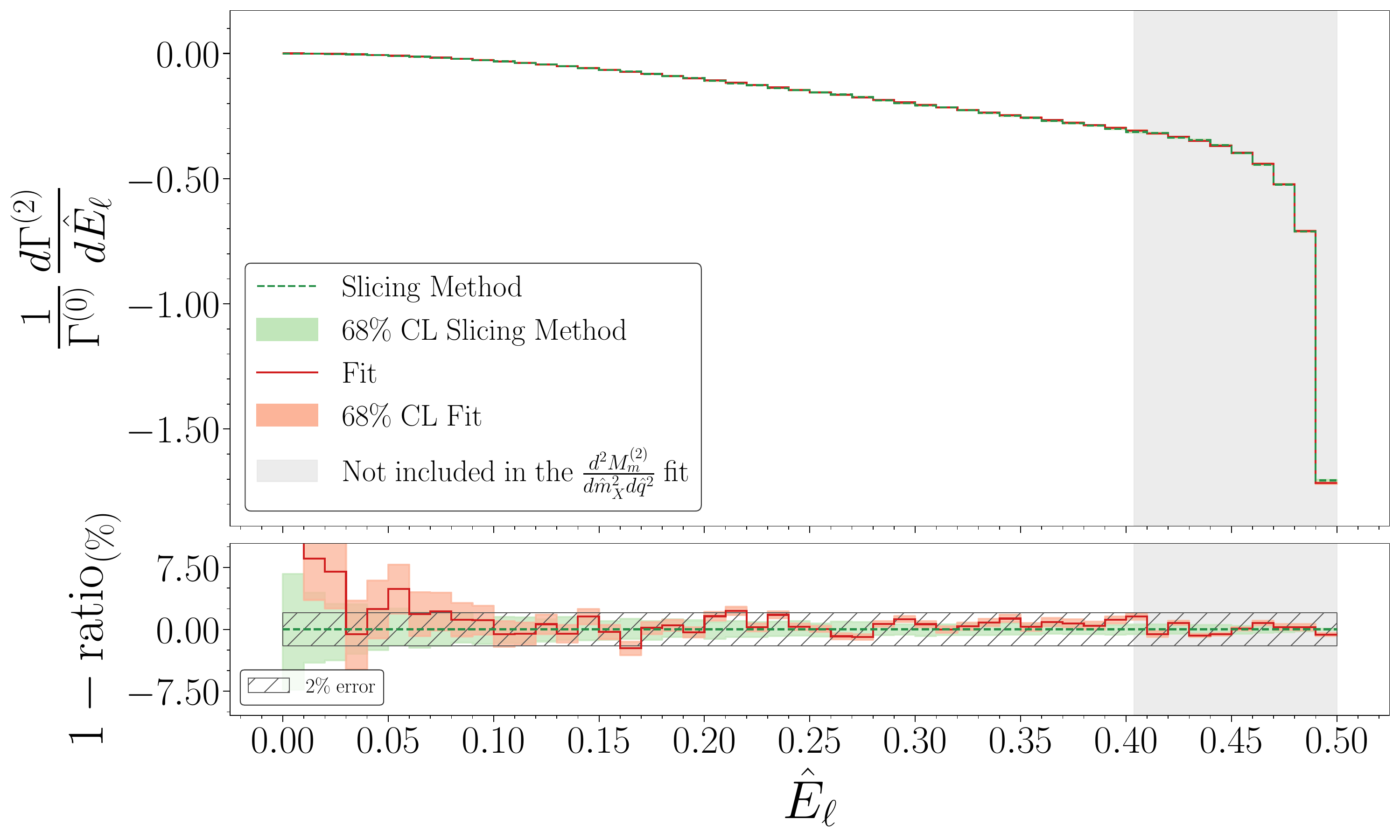}
\\
\includegraphics[width=0.71\textwidth]{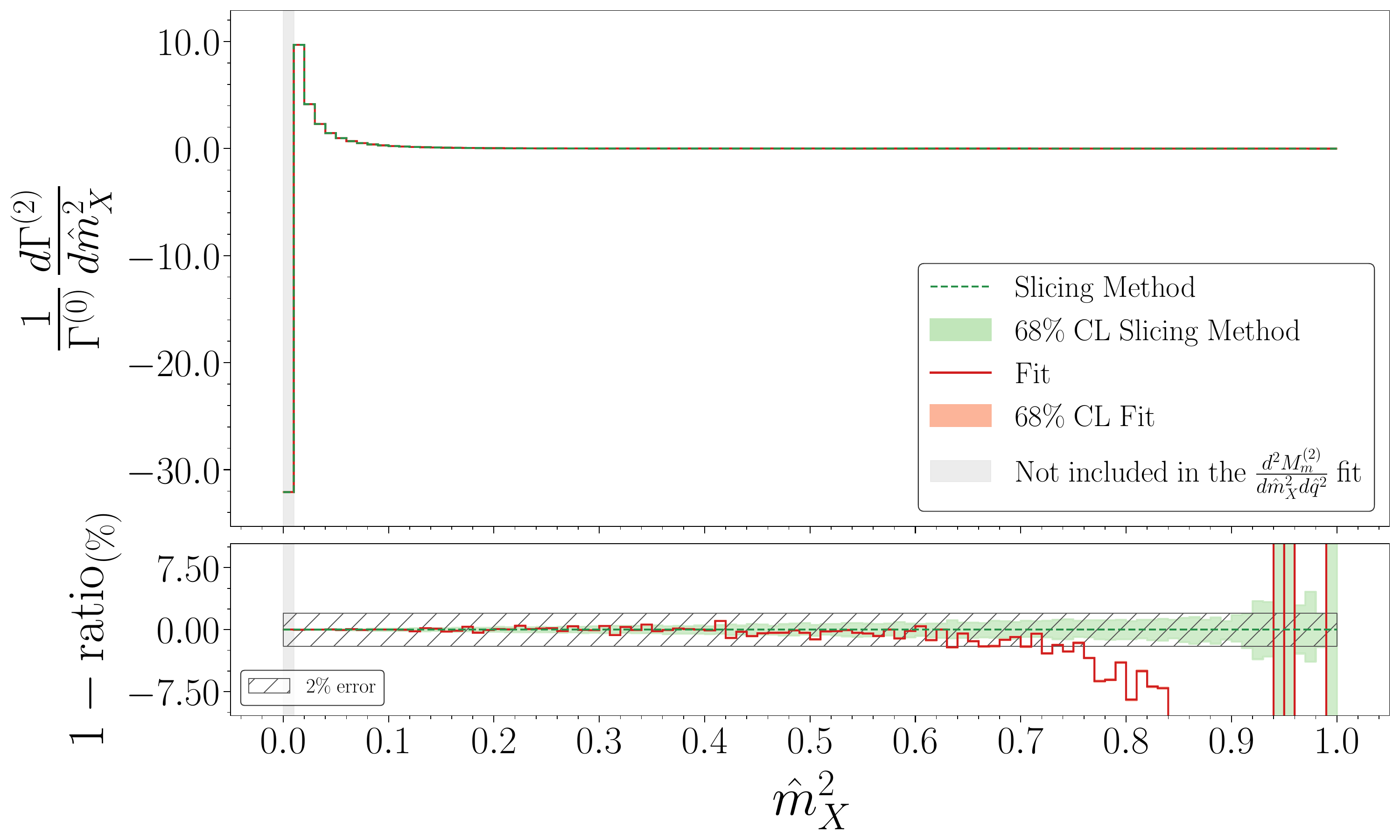}
\\
\includegraphics[width=0.71\textwidth]{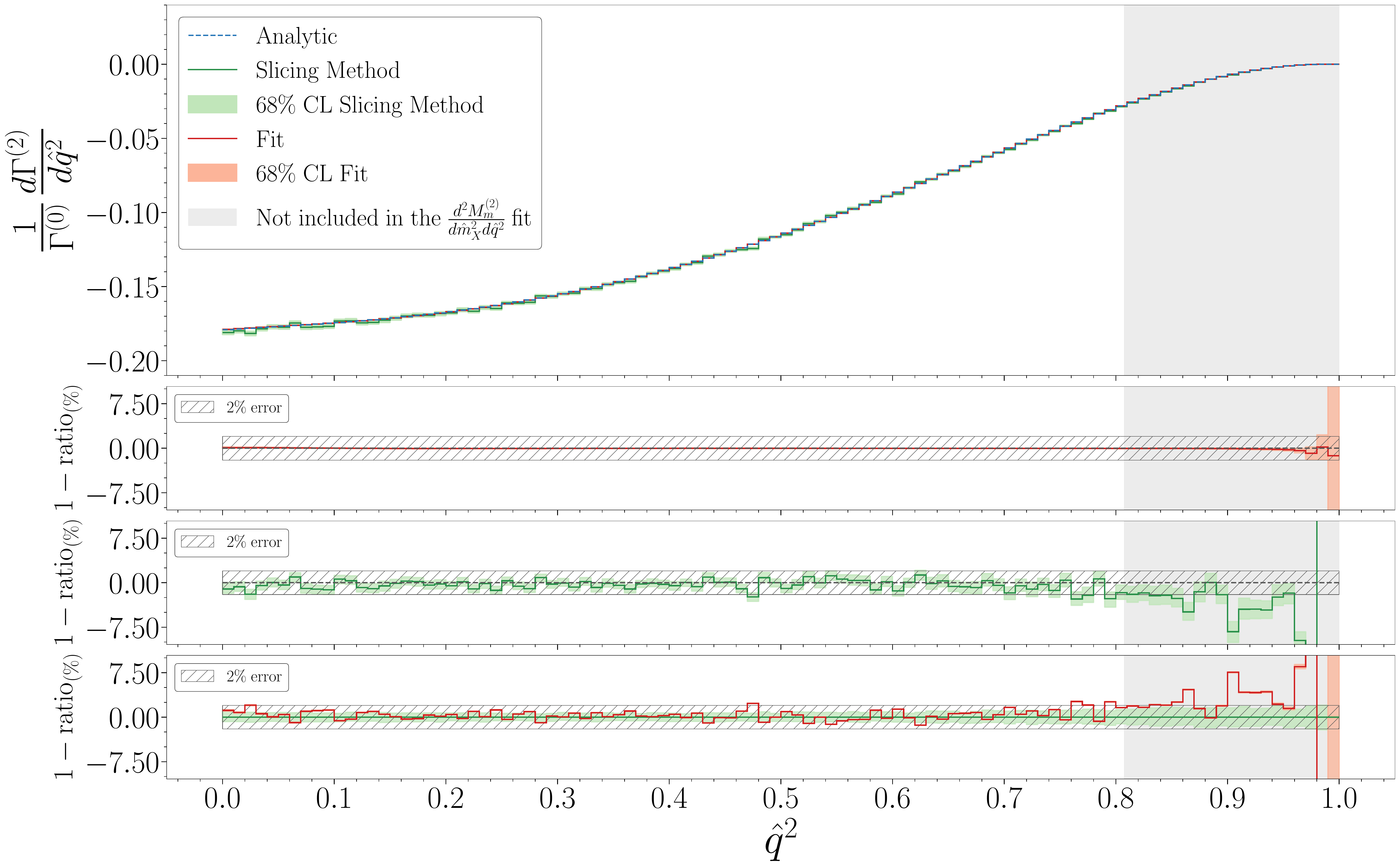}
\caption{NNLO differential decay distributions in $b\to X_u\ell\bar{\nu}$ and comparison with the corresponding results from fitted $W_i^{(2)}$. The fits employed in this Figure are include information from the analytic results for the $\hat{q}^2$ distribution.}
\label{fig:NNLOfitswithq2h}
\end{figure}

\section{Results} 
\label{sec:results}

In this Section we use the semi-analytic expression for the NNLO structure functions $W_i^{(2)}$ that we obtained in this work to evaluate NNLO corrections to several differential distributions, the total decay rate, and energy moments. In addition, we assess the numerical impact of the non-BLM part of the NNLO corrections by comparing them to the BLM NNLO corrections. Unless stated otherwise, all of the results presented in this Section are evaluated by setting $n_l$, the number of light (massless) flavors, equal to 4 and $n_h$, the number of heavy (massive) flavors, equal to 1. The charm quark is considered massless in this framework. In our calculation  $n_l$ can easily be changed by adjusting $\beta_0$ in eq.~(\ref{eq:splitWs}), while only the virtual two-loop corrections depend on $n_h$, which can then be fixed to the desired value. In \cite{Brucherseifer:2013cu}, for instance, the choice was to use $n_h=2,n_l=3$, as this choice appears to approximate well the effect of a massive charm.

\subsection{Differential Distributions}

The differential distributions obtained with the NNLO fits of the form factors $W_i$ are shown in Figures~\ref{fig:NNLOfitswithOUTq2h} and \ref{fig:NNLOfitswithq2h}.

Figure~\ref{fig:NNLOfitswithOUTq2h} shows one-dimensional distributions for the $b\to X_u\ell\bar{\nu}$ process evaluated by using the NNLO fits for the $W_i^{(2)}$ form factors that do \emph{not} include information coming from the analytic expression for the $\hat{q}^2$ distribution. The fits employed in this figure are obtained by using data from the parton level Monte Carlo that evaluates the double differential distribution in $\hat{q}^2$ and $\hat{m}_X^2$, the analytic expression for total decay rate and the analytic results for the first two total lepton energy moments.

The top panel in Figure~\ref{fig:NNLOfitswithOUTq2h} shows the NNLO corrections to the lepton energy spectrum. The red band indicates the result based on the form factors fits; the green band indicates the results obtained directly from the Monte Carlo code. The ratio inlay shows that the agreement between the two calculations is better than 2 \% everywhere, except for very low values of $\hat{E}_l$, where the NNLO corrections  vanish.

The middle panel in Figure~\ref{fig:NNLOfitswithOUTq2h} shows the NNLO corrections to the $\hat{m}_X^2$ distribution. Also in this case, the agreement between the calculation based on the fits and the parton level Monte Carlo is better than 2 \% up to large values of $\hat{m}_X^2$ ($\hat{m}_X^2 \sim 0.75$) where in any case the corrections to this distribution become vanishingly small.

Finally, the bottom panel in Figure~\ref{fig:NNLOfitswithOUTq2h} shows the NNLO corrections to the $\hat{q}^2$ spectrum. In this case, an analytic result is available \cite{Chen:2022wit} and it is shown by the dashed blue line. The three inlays at the bottom show the excellent agreement among the analytic result, the calculation based on the $W_i^{(2)}$ fits and the result obtained from the parton level Monte Carlo. The agreement is better than 2\% except for $\hat{q}^2 > 0.9$, where the NNLO corrections are very small. 

Figure~\ref{fig:NNLOfitswithq2h} shows the same distributions shown in Figure~\ref{fig:NNLOfitswithOUTq2h}, but this time the form factors fits used to obtain the distributions shown by the red bands include information coming from the analytic NNLO corrections to the $\hat{q}^2$ differential distribution.

For all of the three panels in Figure~\ref{fig:NNLOfitswithq2h}, the comments made for the corresponding panels in Figure~\ref{fig:NNLOfitswithOUTq2h} still apply. In particular, the $\hat{E}_l$ and $\hat{m}_X^2$ distributions are affected very little by the inclusion of information about the $\hat{q}^2$ distribution. 

On the contrary, as expected, the use of information about the $\hat{q}^2$ distribution in determining the fit expression for the $W^{(2)}_i$ improves the way in which the distribution obtained from the fits agrees with the analytic NNLO corrections to the $\hat{q}^2$ distribution, in particular for large values of $\hat{q}^2$. 

\subsection{Total Decay Rate and Moments}

\begin{table}[ht]
    \centering
    \begin{tabular}{c|cc}
    & This Work & Comparison\\
      \hline
         $\Gamma^{(2)} / \Gamma^{(0)}$  & $-21.2912 \pm 0.0064$ & $-21.2955$ \cite{vanRitbergen:1999gs}\\
         $L_{1}^{(2)}$   & $-8.1808  \pm 0.0020$ & $-8.1819$  \cite{Pak:2008qt}\\
         $L_{2}^{(2)}$   & $-3.3807  \pm 0.0008$ & $-3.3806$  \cite{Pak:2008qt}\\
         $H_{1}^{(2)}$   & $-6.1274  \pm 0.0037$ & $-6.1150$  \cite{Pak:2008qt}\\
         $H_{2}^{(2)}$   & $-1.8015  \pm 0.0019$ & $-1.7910$  \cite{Pak:2008qt}\\
    \end{tabular}
    \caption{Normalized total decay rate and total moments at NNLO, no cuts on the kinematic variables were applied. Numbers obtained with the fit that does \emph{not} include the $\hat{q}^2$ data. The errors include both the fit uncertainty and the integration error.}
    \label{tab:placeholder}
\end{table}

\begin{table}[ht]
    \centering
    \begin{tabular}{c|cc}
    & This Work & Comparison \\
      \hline
         $\Gamma^{(2)} / \Gamma^{(0)}$  & $-21.2979 \pm 0.0037$ & $-21.2955$ \cite{vanRitbergen:1999gs}\\
         $L_{1}^{(2)}$   & $-8.1822  \pm 0.0013$ & $-8.1819$  \cite{Pak:2008qt}\\
         $L_{2}^{(2)}$   & $-3.3812  \pm 0.0006$ & $-3.3806$  \cite{Pak:2008qt}\\
         $H_{1}^{(2)}$   & $-6.1133  \pm 0.0018$ & $-6.1150$  \cite{Pak:2008qt}\\
         $H_{2}^{(2)}$   & $-1.7893  \pm 0.0009$ & $-1.7910$  \cite{Pak:2008qt}\\
    \end{tabular}
    \caption{Normalized total decay rate and total moments at NNLO, no cuts on the kinematic variables were applied. Numbers obtained with the fit that includes the $\hat{q}^2$ data. The  errors include both the fit uncertainty and the integration error.}
    \label{tab:placeholderwithq2h}
\end{table}

As an additional check of our calculations, we study how the fits of the $W_i$ form factors reproduce known analytic results for the NNLO corrections to the total decay rate and to the total leptonic and hadronic energy moments. The NNLO corrections to the total $b\to X_u\ell\bar{\nu}$ decay rate have been known for a long time \cite{vanRitbergen:1999gs}. Table~\ref{tab:placeholder} shows that integrating eq.~\eqref{eq:double} with the $W^{(2)}_i$ obtained through the fitting procedure described in this work leads to a NNLO correction that shows a four digits agreement with the analytic result. 

Similarly, the corrections to the total leptonic and hadronic energy moments can be defined through the perturbative expansions
\begin{equation}
    \frac{1}{\Gamma^{(0)}} \langle E^n_\ell \rangle =  \sum_{j = 0}^\infty \left( \frac{\alpha_s}{\pi} \right)^j L^{(j)}_n \, , \qquad
    \frac{1}{\Gamma^{(0)}} \langle E^n_X \rangle =  \sum_{j = 0}^\infty \left( \frac{\alpha_s}{\pi} \right)^j H^{(j)}_n \, ,
\end{equation}
where $E_\ell$ and $E_X$ are the energies of the charged lepton and of the hadronic final state, respectively. These quantities were evaluated in \cite{Pak:2008qt}. Again, the calculation of the integrated moments based on the form factor fits leads to NNLO corrections that have two-three  digits of agreement with the exact results. The agreement of  $H^{(2)}_1$ and $H^{(2)}_2$ obtained from our approximate structure functions with the corresponding analytic values found in \cite{Pak:2008qt} is better in Table~\ref{tab:placeholderwithq2h} than in Table~\ref{tab:placeholder}: the inclusion of information coming from the $\hat{q}^2$ distribution in the fits has a tangible effect in the calculation of these quantities. In Table~\ref{tab:placeholderwithq2h} the agreement is better than 0.1\%.

We have also computed the moments subject to the kinematical cuts employed in \cite{Brucherseifer:2013cu}, setting $n_h=2$ and $n_l=3$. Our results show excellent agreement with Table 1 of \cite{Brucherseifer:2013cu}, generally within their stated uncertainty, with maximal deviations of  1.3\%.

\subsection{Impact of the NNLO Structure Functions}
 
\begin{figure}[htp]
\begin{center}
\includegraphics[width=0.71\textwidth]{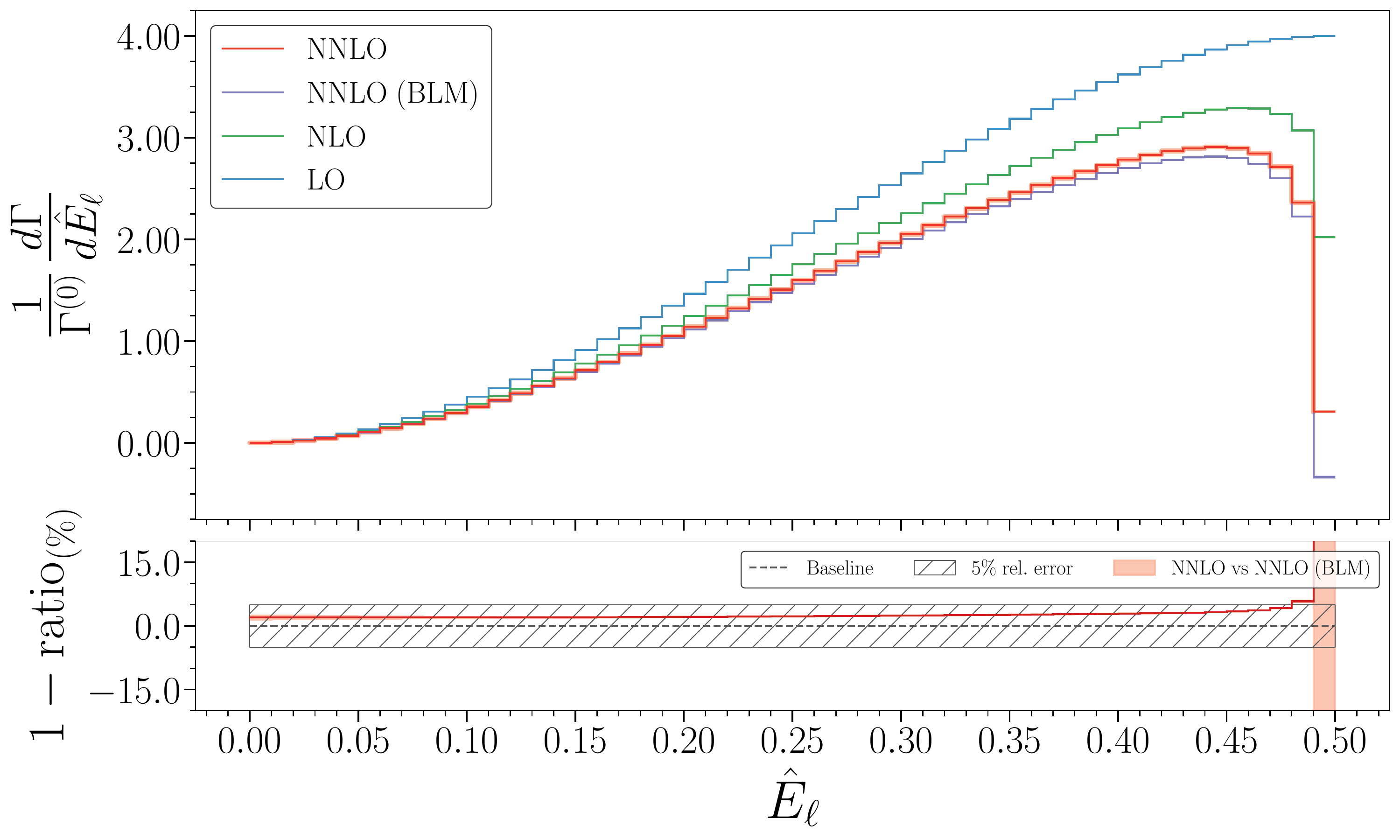}
\\
\includegraphics[width=0.71\textwidth]{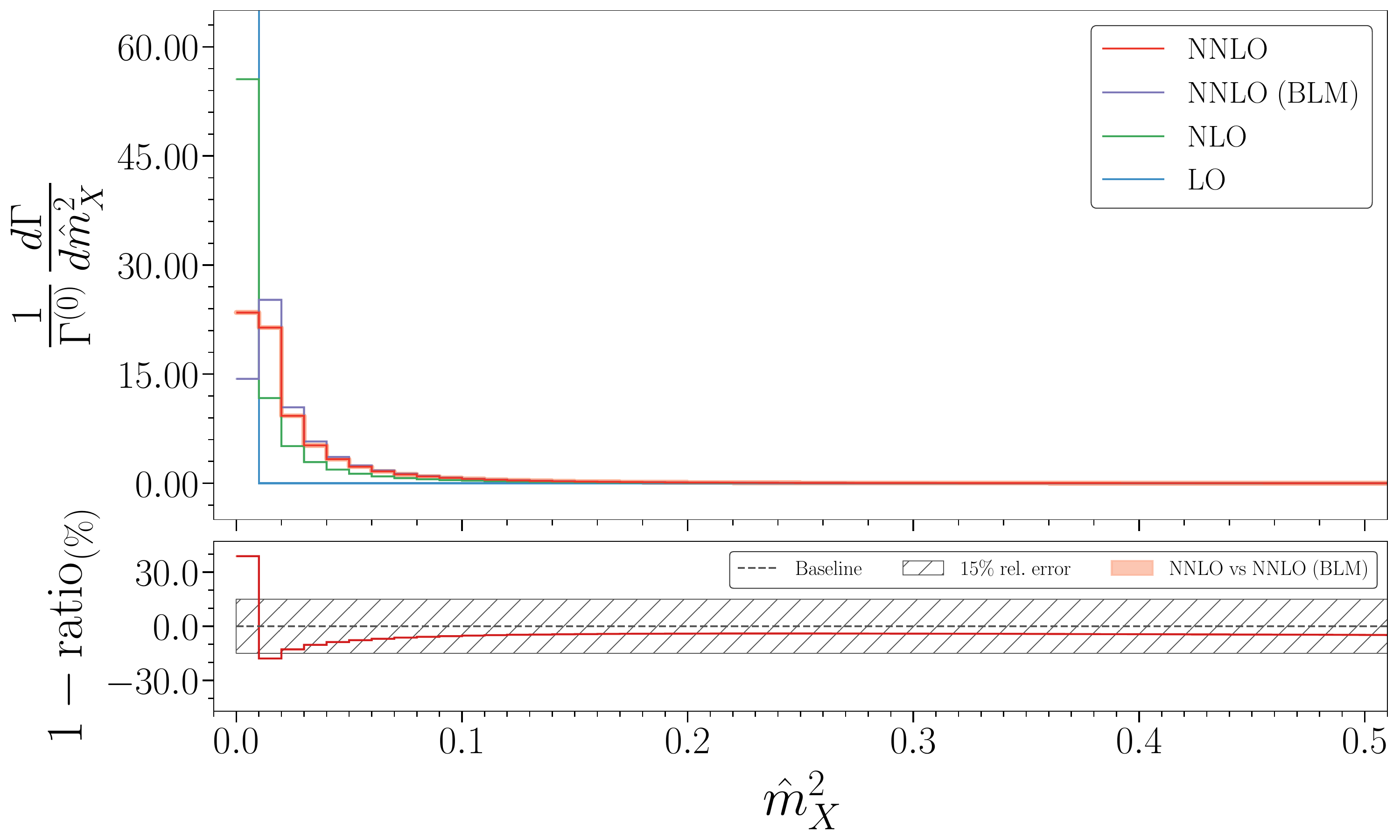}
\\
\includegraphics[width=0.71\textwidth]{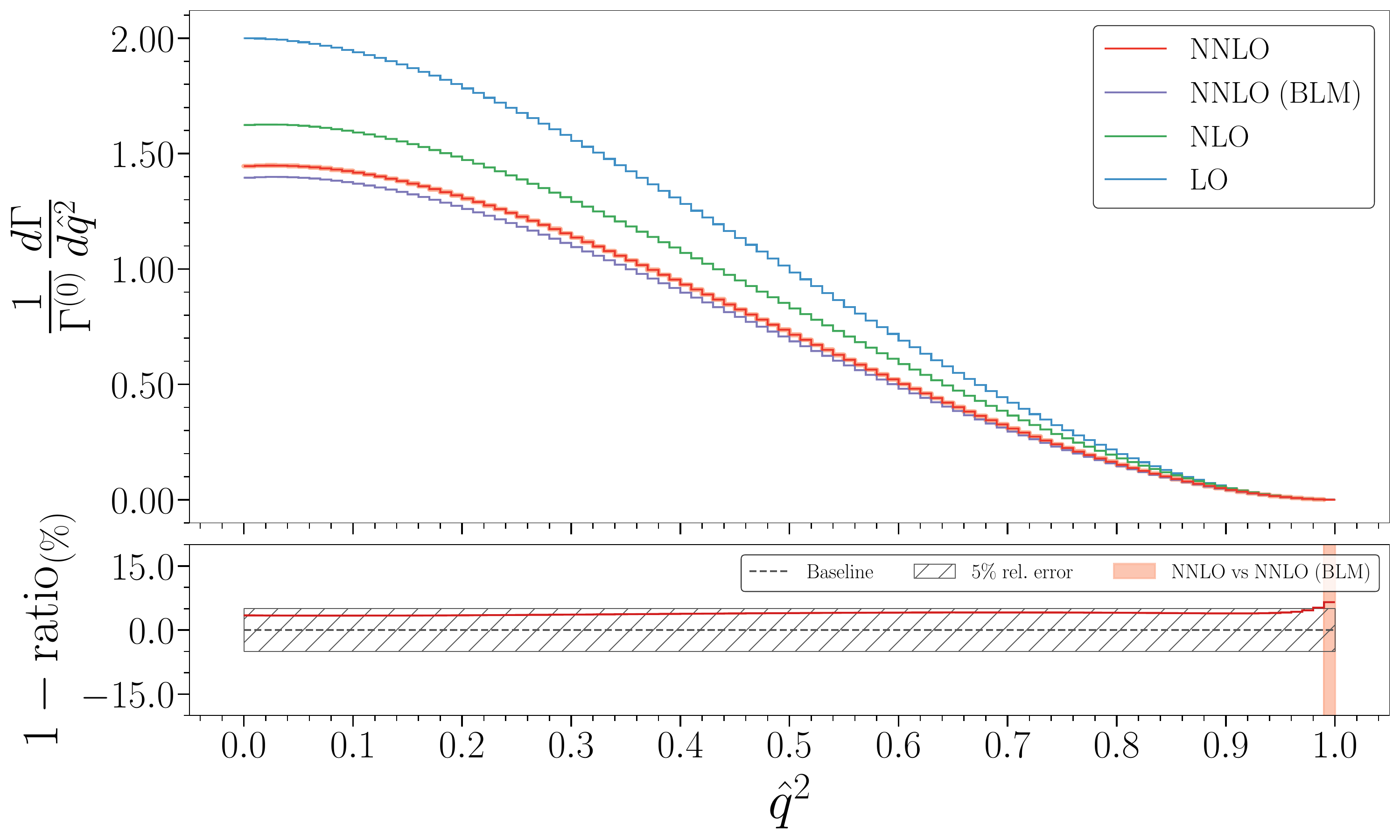}
\caption{Comparison of the different perturbative orders to the one-dimensional differential distributions.
The NNLO contributions are evaluated by using the fitted form factors obtained by including data coming from the $\hat{q}^2$ distribution. }
\label{fig:differentOrders}
\end{center}
\end{figure}

Figure~\ref{fig:differentOrders} shows the impact of the various perturbative orders on the differential spectra with 
respect to $\hat{E}_\ell$, $\mxs$, and $\hat{q}^2$. In particular, the goal here is to understand what is the effect of the NNLO corrections that are not captured by the BLM approximation. The NNLO curves in Figure~\ref{fig:differentOrders} refer to differential distributions calculated starting from form factors fitted by including also data from the $\hat{q}^2$ distribution. We observe that in all three distributions the non-BLM NNLO corrections lead to a relatively flat shift of the NLO result everywhere in the allowed range, except near the kinematic endpoints (i.e.~$\hat{E}_\ell \sim 0.5$, $\mxs \sim 0$, $\hat{q}^2 \sim 1$), where the effect of the non-BLM NNLO corrections is more noticeable.  In particular, the inlay in each panel shows that the non-BLM NNLO corrections, away from the endpoint, amount to $2-3 \%$ in the $\hat{E}_\ell$ distribution, $4-6\%$ in the $\mxs$ distribution and $3-4 \%$ in the $\hat{q}^2$ distribution. The ratio between the non-BLM and the BLM NNLO corrections is generally around  -20\% to -30\%, depending on the kinematic distribution,  in agreement with  \cite{Brucherseifer:2013cu}.

\section{Summary}
\label{sec:conclusions}

In this paper, we have computed  the perturbative NNLO contributions to the structure functions $W_i$ for the decay $b\to X_u\ell\bar{\nu}$ (equivalently, the triple-differential distribution) using a semi-analytic method. Our results combine the exact singularity structure implied by HQET+SCET factorization formulas, the full BLM corrections computed analytically in Ref.~\cite{Gambino:2006wk}, and a numerical calculation performed with a parton-level Monte Carlo code. We further constrain the fit that yields our final expressions by including several additional analytic results: the total width \cite{vanRitbergen:1999gs}, the first two leptonic moments \cite{Czarnecki:2001cz, Pak:2008qt}, and the $q^2$-distribution \cite{Chen:2022wit}.

The resulting formulas provide accurate approximations to the $O(\alpha_s^2)$ contributions to the structure functions and are ready for phenomenological applications. They reproduce our numerical results to better than 2\% (and typically significantly better) and are validated against a range of available analytic and numerical benchmarks. All results are supplied as ancillary files in both Mathematica and Python formats. These expressions enable an improved description of inclusive $B\to X_u\ell\bar{\nu}$ decays and support a more precise determination of $|V_{ub}|$.

\subsection*{Acknowledgments}
We thank Ben Pecjak for providing the SCET hard functions in analytic form. We thank Yitao Li for his work in the analysis of the output of the parton level Monte Carlo code. A.B. would like to thank Simone Alioli and Thomas Mannel for discussions. We thank Matthew A. Lim for his collaboration during the early stages of this work. B.C. would like to thank the Phenomenology Group at DAMTP for useful discussions. P.G. is supported in part by the Italian Ministry of University and Research (MUR) and the European Union (EU) – Next Generation EU, Mission 4, Component 1, PRIN 2022, grant 2022N4W8WR, CUP D53D23002830006. The work of B.C. is supported by the La Caixa Junior Leader fellowship from the “la Caixa” Foundation (ID 100010434, fellowship code LCF/BQ/PI24/12040024). The work of A.F. si supported in part by the PSC Cycle 56 award 68092-00 56.

\newpage

\appendix

\section{Coefficients for the NNLO Fit: Numerical Values}

In this appendix we list the numerical values of the coefficients of our final NNLO fit.

\vspace{0.25cm}

\begin{tabular}[H]{|c|c|c|c|}%
\hline%
 &$i = 1$&$i = 2$&$i = 3$\\%
\hline%
$\alpha_{i1}^{(2)}$&4952.7187&{-}31407.1332&5517.3024\\[2pt]%
$\alpha_{i2}^{(2)}$&{-}3947.7360&136163.9853&52851.5819\\[2pt]%
$\alpha_{i3}^{(2)}$&562.4255&{-}1256.7592&635.2231\\[2pt]%
$\alpha_{i4}^{(2)}$&31.1793&3952.0358&1413.4776\\[2pt]%
$\alpha_{i5}^{(2)}$&1770.0165&{-}115430.1013&{-}55144.9066\\[2pt]%
$\alpha_{i6}^{(2)}$&{-}1206.9890&52675.0602&27112.1379\\[2pt]%
$\alpha_{i7}^{(2)}$&308.8605&{-}9923.5029&{-}5682.9691\\[2pt]%
$\alpha_{i8}^{(2)}$&{-}77.3144&896.4481&597.9928\\[2pt]%
$\alpha_{i9}^{(2)}$&5792.1777&{-}69794.8722&{-}26394.6664\\[2pt]%
$\alpha_{i10}^{(2)}$&{-}531.0761&14584.0834&5675.9926\\[2pt]%
$\alpha_{i11}^{(2)}$&{-}50.6792&{-}2024.0984&{-}703.9054\\[2pt]%
$\alpha_{i12}^{(2)}$&{-}727.9694&9689.3386&4062.1451\\[2pt]%
$\alpha_{i13}^{(2)}$&60.5978&{-}2279.3031&{-}1476.9621\\[2pt]%
$\alpha_{i14}^{(2)}$&21.6398&22.4885&{-}83.4749\\[2pt]%
$\alpha_{i15}^{(2)}$&12277.6207&{-}148952.8709&{-}56241.7233\\[2pt]%
$\alpha_{i16}^{(2)}$&136601.0409&{-}38580684.9366&{-}121150203.3360\\[2pt]%
$\alpha_{i17}^{(2)}$&8147.8085&{-}10284462.7456&{-}26464028.9372\\[2pt]%
$\alpha_{i18}^{(2)}$&{-}6751.4580&146777.3140&49599.6348\\[2pt]%
$\alpha_{i19}^{(2)}$&0.9835&8.0918&18.6702\\[2pt]%
$\alpha_{i20}^{(2)}$&{-}8.8579&3994.7380&1364.7214\\[2pt]%
$\alpha_{i21}^{(2)}$&125514.9713&{-}38489842.9302&{-}121111667.5023\\[2pt]%
$\alpha_{i22}^{(2)}$&{-}219207.3143&75667151.1376&225177336.1119\\[2pt]%
$\alpha_{i23}^{(2)}$&4216.5134&7596778.0723&17793979.4498\\[2pt]%
$\alpha_{i24}^{(2)}$&13422.3881&{-}157307.5548&{-}58874.0411\\[2pt]%
$\alpha_{i25}^{(2)}$&{-}0.7973&176.1441&{-}81.8281\\[2pt]%
$\alpha_{i26}^{(2)}$&9.0049&{-}5492.8915&{-}1918.5180\\[2pt]%
$\alpha_{i27}^{(2)}$&{-}159617.3532&66654868.1612&191071639.7906\\[2pt]%
$\alpha_{i28}^{(2)}$&76173.4911&{-}27661708.6088&{-}80855776.4959\\[2pt]%
$\alpha_{i29}^{(2)}$&{-}12861.3939&3580224.7838&10840357.6808\\[2pt]%
$\alpha_{i30}^{(2)}$&{-}6658.1514&21534.6674&12762.3888\\[2pt]%
$\alpha_{i31}^{(2)}$&{-}0.5036&57.8697&83.0491\\[2pt]%
$\alpha_{i32}^{(2)}$&{-}0.5898&1635.4651&568.2341\\%
\hline%
\end{tabular}%

\section{Appendix: Python and Mathematica Packages}

The ancillary material includes a \texttt{Python} package (\texttt{buW2}) that provides the ingredients needed to evaluate and integrate the $b\to X_u\ell\bar{\nu}$ structure functions $W_i$ up to order NNLO, including our fitted NNLO real-radiation contributions. At a given perturbative order, the form factors can be written schematically as
\begin{align}
W_i^{(n)}(\hat{q}_0, \hat{q}^2) &= W_i^{(n, \delta)}(\hat{q}^2)\, \delta(1 + \hat{q}^2 - 2\hat{q}_0)\nonumber\\
&+ \sum_{m = 0}^{2 n -1} W_i^{(n, +)}(\hat{q}_0, \hat{q}^2) \left[\dfrac{\ln^m(1 + \hat{q}^2 - 2\hat{q}_0)}{1 + \hat{q}^2 - 2\hat{q}_0}\right]_{+} + R_i^{(n)}(\hat{q}_0, \hat{q}^2) .
\end{align}
For what concerns the evaluation of NNLO structure functions, within our approach, the $R_i^{(n)}$ functions will be combinations of basis functions, with fitted coefficients, as in eq.~\eqref{eq:R_i_basis of functions}.

For a generic probe $f(\hat{q}_0,\hat{q}^2)$ and an integration range
\begin{equation}
\hat{q}_0^\text{min} \leq \hat{q}_0\leq \frac{1 + \hat{q}^2}{2}\, ,
\end{equation}
the package organises the $\hat{q}_0$ integral into virtual, real, subtraction, and surface contributions:
\begin{align}
&\int_{\frac{\hat{q}^2}{2}}^{\frac{1 + \hat{q}^2}{2}}d\hat{q}_0\, f(\hat{q}_0, \hat{q}^2)\,W_i^{(n)}(\hat{q}_0, \hat{q}^2) = \nonumber\\[2pt]
& \dfrac{1}{2}\,f\left(\frac{1 + \hat{q}^2}{2}, \hat{q}^2\right)\, W_i^{(n, \delta)}(\hat{q}^2)\nonumber\\[2pt]
&+ \int_{\hat{q}_0^\text{min}}^{\frac{1 + \hat{q}^2}{2}}d\hat{q}_0\, f(\hat{q}_0, \hat{q}^2)\left(\sum_{m = 0}^{2n-1} W_i^{(n, +)}(\hat{q}_0, \hat{q}^2) \dfrac{\ln^m(1 + \hat{q}^2 - 2\hat{q}_0)}{1 + \hat{q}^2 - 2\hat{q}_0} + R_i^{(n)}(\hat{q}_0, \hat{q}^2)\right)\nonumber\\[2pt]
&+ \int_{\hat{q}_0^\text{min}}^{\frac{1 + \hat{q}^2}{2}}d\hat{q}_0\, f\left(\frac{1 + \hat{q}^2}{2}, \hat{q}^2\right)\left(-\sum_{m = 0}^{2n-1} W_i^{(n, +)}\left(\frac{1 + \hat{q}^2}{2}, \hat{q}^2\right) \dfrac{\ln^m(1 + \hat{q}^2 - 2\hat{q}_0)}{1 + \hat{q}^2 - 2\hat{q}_0}\right)\nonumber\\[2pt]
&+ f\left(\frac{1 + \hat{q}^2}{2}, \hat{q}^2\right)\sum_{m = 0}^{2n-1} W_i^{(n, \hat{q}_0^\text{min})}(\hat{q}^2)\,\dfrac{\ln^{m + 1}(1 + \hat{q}^2 - 2\hat{q}_0^\text{min})}{2(m + 1)}\, .
\end{align}
The terms $W_i^{(n,\hat{q}_0^\text{min})}$ accounts for surface contributions that appear when the plus-distributions,
\begin{equation}
\left[\dfrac{\ln^m(1 + \hat{q}^2 - 2\hat{q}_0)}{1 + \hat{q}^2 - 2\hat{q}_0}\right]_{+}\, , \nonumber
\end{equation}
are integrated with a lower limit different from their defining one,
\begin{align}
\int d\hat{q}_0\, f(\hat{q}_0, \hat{q}^2)\left[\dfrac{\ln^m(1 + \hat{q}^2 - 2\hat{q}_0)}{1 + \hat{q}^2 - 2\hat{q}_0}\right]_{+} = \int_{\frac{\hat{q}^2}{2}}^{\frac{1 + \hat{q}^2}{2}}d\hat{q}_0\,& \dfrac{\ln^m(1 + \hat{q}^2 - 2\hat{q}_0)}{1 + \hat{q}^2 - 2\hat{q}_0} \nonumber\\
& \times \left[f(\hat{q}_0, \hat{q}^2) - f\left(\frac{1 + \hat{q}^2}{2}, \hat{q}^2\right)\right] \, .
\end{align}
Note that, by construction of the sufrace terms coming from the plus-distribution prescription, one has that $W_i^{(n, \hat{q}_0^\text{min})}(\hat{q}^2) = W_i^{(n, +)}\left(\dfrac{1 + \hat{q}}{2}, \hat{q}^2\right)$.

These pieces are implemented in the package via the following classes:
\begin{verbatim}
w0, wV1, wR1, wR1Subt, wR1Surf, wV2, wR2, wR2Subt, wR2Surf
\end{verbatim}
The correspondence is
\begin{align}
    W_i^{(n, \delta)}(\hat{q}^2) \longmapsto &\; \texttt{w0},\, \texttt{wV1},\, \texttt{wV2} \, , \nonumber \\
    \sum_{m = 0}^{M_n} W_i^{(n, +)}(\hat{q}_0, \hat{q}^2) \dfrac{\ln^m(1 + \hat{q}^2 - 2\hat{q}_0)}{1 + \hat{q}^2 - 2\hat{q}_0} + R_i^{(n)}(\hat{q}_0, \hat{q}^2) \longmapsto &\; \texttt{wR1},\, \texttt{wR2} \,  , \nonumber \\
    -\sum_{m = 0}^{M_n} W_i^{(n, +)}\left(\frac{1 + \hat{q}^2}{2}, \hat{q}^2\right) \dfrac{\ln^m(1 + \hat{q}^2 - 2\hat{q}_0)}{1 + \hat{q}^2 - 2\hat{q}_0} \longmapsto &\; \texttt{wR1Subt},\, \texttt{wR2Subt}\,, \nonumber \\
    \sum_{m = 0}^{M_n} W_i^{(n, \hat{q}_0^\text{min})}(\hat{q}^2)\,\dfrac{\ln^m(1 + \hat{q}^2 - 2\hat{q}_0^\text{min})}{2(m + 1)} \longmapsto &\; \texttt{wR1Surf},\, \texttt{wR2Surf}\, .\label{eq:W_i_q0h_int_probe}
\end{align}
That is, \texttt{w0} provides the leading-order contribution to the structure functions, \texttt{wV1} and \texttt{wV2} the virtual terms, \texttt{wR1} and \texttt{wR2} the real-radiation contribution (singular and regular), \texttt{wR1Subt} and \texttt{wR2Subt} the subtraction implied by the plus-distribution prescription, and \texttt{wR1Surf} and \texttt{wR2Surf} the associated surface terms.

After installing the dependencies and setting the \texttt{conda} environment as described in \texttt{README.md}, the basic usage is:
\begin{verbatim}
import jax
import jax.numpy as jnp
from jax.config import config;
config.update("jax_enable_x64", True)

import yaml

import buW2
\end{verbatim}
Load the input parameters (\texttt{buW2/parameters.yaml}) and fit coefficients\\(\texttt{buW2/wi\_R2\_parameters.yaml}):
\begin{verbatim}
with open("buW2/parameters.yaml", "r") as input_file:
    pars = yaml.load(input_file, Loader = yaml.FullLoader)

pars = pars["Parameters"]

with open("buW2/wi_R2_parameters.yaml", "r") as input_file:
    wi_R2_pars = yaml.load(input_file, Loader = yaml.FullLoader)

wi_R2_pars = jnp.array(wi_R2_pars["pars"])
\end{verbatim}
Instantiate the classes,
\begin{verbatim}
wv2 = buW2.wV2(buW_pars = pars)
wr2 = buW2.wR2(buW_pars = pars)
wr2_subt = buW2.wR2Subt(buW_pars = pars)
wr2_surf = buW2.wR2Surf(buW_pars = pars)
\end{verbatim}
and evaluate them, e.g.\ on vector inputs:
\begin{verbatim}
eps = 1e-3

q2h_vec = jnp.linspace(0.15, 0.35, 50)

q2h_tensor = jnp.tile(q2h_vec[None], reps = (25, 1))
q0h_tensor = jnp.linspace(
    jnp.sqrt(q2h_vec) + eps, ((1 + q2h_vec) / 2) - eps, 25
)

q0h_min = 0.10

w1_v2_tensor = wv2("W1", q2h_tensor)
w2_v2_tensor = wv2("W2", q2h_tensor)
w3_v2_tensor = wv2("W3", q2h_tensor)

w1_r2_tensor = wr2("W1", q0h_tensor, q2h_tensor, wi_R2_pars)
w2_r2_tensor = wr2("W2", q0h_tensor, q2h_tensor, wi_R2_pars)
w3_r2_tensor = wr2("W3", q0h_tensor, q2h_tensor, wi_R2_pars)

w1_r2_subt_tensor = wr2_subt("W1", q0h_tensor, q2h_tensor)
w2_r2_subt_tensor = wr2_subt("W2", q0h_tensor, q2h_tensor)
w3_r2_subt_tensor = wr2_subt("W3", q0h_tensor, q2h_tensor)

w1_r2_surf_tensor = wr2_surf("W1", q2h_tensor, q0h_min)
w2_r2_surf_tensor = wr2_surf("W2", q2h_tensor, q0h_min)
w3_r2_surf_tensor = wr2_surf("W3", q2h_tensor, q0h_min)
\end{verbatim}
All routines accept either scalars or \texttt{JAX} arrays and are vectorised over input shapes.

Finally, the ancillary files also include a \texttt{jupyter} notebook demonstrating the evaluation of each class and illustrating the numerical integration of the form factors against a smooth probe function, as in eq.~\eqref{eq:W_i_q0h_int_probe}.

Note that in \texttt{buW2} we use an approximation for the polylogs involved in the computations. This can lead to small numerical difference with respect to the \texttt{Mathematica} expressions when evaluated with the HPL package \cite{Maitre:2005uu, Maitre:2007kp}.

The structure of the \texttt{Mathematica} file is self explanatory; the file loads the HPL package \cite{Maitre:2005uu, Maitre:2007kp} which is needed for the evaluation of the functions that appear in the double virtual corrections.
The \texttt{Mathematica} file loads the LO, NLO and NNLO results for the three structure functions $W_i$ as defined in eq.~\eqref{eq:triple}. Once the values for $\texttt{q2h}\equiv\hat{q}^2$ and $\texttt{q0h}\equiv\hat{q}_0$ are given, the functions \texttt{W1[q2h,q0h]}, \texttt{W2[q2h,q0h]}, \texttt{W3[q2h,q0h]} provide the numerical coefficients of the Dirac delta function, the plus distributions and the fitted regular remainder up to NNLO.

\bibliography{biblio}
\bibliographystyle{JHEP}

\end{document}